\documentclass[letterpaper]{aa}
\usepackage{graphicx}
\usepackage{natbib}
\bibpunct{(}{)}{;}{a}{}{,}    
\newcommand{\be}{\begin{equation}}
\newcommand{\ee}{\end{equation}}
\newcommand{\beq}{\begin{equation}}
\newcommand{\eeq}{\end{equation}}
\newcommand{\bea}{\begin{eqnarray}}
\newcommand{\eea}{\end{eqnarray}}

\newcommand{\subscr}[1]{_\mathrm{#1}}

\newcommand{\gsim}{\raisebox{-0.6ex}{$\stackrel{{\displaystyle>}}{\sim}$}}

\newcommand{\url}[1]{{\tt #1}}
\newcommand{\pomega}{{\varpi}}

\newcommand{\doverdt}[1]{\frac{\partial #1}{\partial t}}
\newcommand{\qonex}{1.75}
\newcommand{\qtwox}{5.9}
\def\gapp{\lower 3pt\hbox{${\buildrel > \over \sim}$}\ }
\def\lapp{\lower 3pt\hbox{${\buildrel < \over \sim}$}\ }
\def\Mjup{{\rm M\subscr{Jup}}}
\def\Msol{{\rm M_\odot}}

\def\degx{{^\circ}}
\def\GJ{{\rm GJ~876}}

\newcommand{\yr}{{\rm\,yr}}
\newcommand{\au}{{\rm\,AU}}
\begin{document}
\title{Modeling the resonant planetary system GJ~876}
\author{
Wilhelm Kley\inst{1}
\and
Man Hoi Lee\inst{2}
\and
Norman Murray\inst{3}
\and
Stanton J. Peale\inst{2}}
\offprints{W. Kley,\\ \email{kley@tat.physik.uni-tuebingen.de}}
\institute{
     Institut f\"ur Astronomie \& Astrophysik, 
     Abt. Computational Physics,
     Universit\"at T\"ubingen,
     Auf der Morgenstelle 10, D-72076 T\"ubingen, Germany
\and
Department of Physics, University of California, Santa Barbara, CA 93106, USA
\and
Canada Research Chair in Astrophysics; 
CITA, University of Toronto, 60 St. George Street Toronto, Ontario M5S 3H8, Canada
}
\date{Received: January 7, 2005 / Accepted: March 24, 2005}
\abstract{
The two planets about the star GJ 876 appear to have undergone
extensive migration from their point of origin in the protoplanetary
disk --- both because of their close proximity to the star (30 and 60
day orbital periods) and because of their occupying three stable
orbital resonances at the 2:1 mean-motion commensurability.
The resonances were most likely established by converging differential
migration of the planets leading to capture into the resonances.
A problem with this scenario is that continued migration of the system
while it is trapped in the resonances leads to orbital eccentricities
that rapidly exceed the observational upper limits of $e_1 \approx
0.31$ and $e_2 \approx 0.05$.
As seen in forced 3-body simulations, these lower eccentricities would
persist during migration only for an eccentricity damping rate $\dot
e_2/e_2$ exceeding $\approx 40\, \dot a_2/a_2$.
Previous theoretical and numerical analyses have found $\dot e/e \sim
\dot a/a$ or even eccentricity growth through disk-planet
interactions.

In an attempt to find effects that could relax the excessive
eccentricity damping requirement, we explore the evolution of the GJ
876 system using two-dimensional hydrodynamical simulations that
include viscous heating and radiative cooling in some cases.
Before we evolve the whole system, the disk with just the outer planet
embedded is brought into equilibrium. We find that the relaxed disk
remains circular in all models for low planet-to-star mass ratios
$q_2$, but becomes eccentric for high mass ratios for those models
with fixed temperature structure. The disk in models with full
radiative thermodynamics remains circular for all $q_2$ considered
due to the larger disk temperatures.
Given the small stellar mass, the mass ratio for the GJ 876 system is
rather high (with minimum $q_2 = 5.65 \times 10^{-3}$), and so the GJ
876 disk may have been slightly eccentric during the migration.

With a range of parameter values, we find that a hydrodynamic
evolution within the resonance, where only the outer planet interacts
with the disk, always rapidly leads to large values of eccentricities
that exceed those observed --- similar to the three-body results.
The resonance corresponding to the resonant angle $\theta_1 =
2\lambda_2 - \lambda_1 - \varpi_1$ (involving the inner planet's
periapse longitude, $\varpi_1$) is always captured first.
There is no additional delay in capturing $\theta_2 = 2\lambda_2 -
\lambda_1 - \varpi_2 $ into resonance that is
attributable to the secular prograde contribution to the precession of
$\varpi_2$ from the interaction with the disk, but an eccentric disk
can induce a large outer planet eccentricity $e_2$ before capture and
thereby further delay capture of $\theta_2$ for larger planetary
masses.  The delay in capturing $\theta_2$ into libration, while
delaying the resonance-induced growth of $e_2$, has no effect on the
forced eccentricities of both 
planets, which are uniquely determined by the resonance conditions,
once both $\theta_j$ are librating.  

Only if mass is removed from the disk on a time scale of the order
of the migration time scale (before
there has been extensive migration after capture), as might occur for
photoevaporation in the late phases of planet formation, can we end up with
eccentricities that are consistent with the observations. 

\keywords{accretion disks -- 
          planet formation --
          hydrodynamics -- 
          celestial mechanics
          }}
\maketitle
\markboth
{Kley, et al.: Modeling GJ~876}
{Kley, et al.: Modeling GJ~876}
\section{Introduction}
\label{sec:introduction}
Among the 14 known extrasolar planetary systems with multiple planets,
at least three exhibit pairs of planets that are likely to be in orbital
resonances at low order commensurabilities of their mean motions. The
pair of planets around GJ 876 \citep{2001ApJ...556..296M} is well
confirmed to be deep in the resonances associated with the 2:1 mean
motion commensurability \citep{2001ApJ...551L.109L, 
2001ApJ...558..392R, 2004astro-ph0407441}, 
where resonant angles $\theta_1 = 2\lambda_2 - \lambda_1 - \varpi_1$,
$\theta_2 = 2\lambda_2 - \lambda_1 - \varpi_2$, and $\Delta\varpi =
\varpi_2 - \varpi_1=\theta_1-\theta_2$  are all librating
 about $0^\circ$ with small
amplitudes. Here $\lambda_j$ are mean longitudes and $\varpi_j$ are
longitudes of periapse, both numbered from the inside out.  The pair of
planets around HD 82943 is likely to also be in resonances at 2:1
\citep{2004A&A...415..391M}, and the middle pair of planets in the 4
planet system orbiting 55 Cnc may be in resonances at 3:1 
\citep{2002ApJ...581.1375M,2004ApJ...614L..81M}. Thus as many as one-fourth of
multiple-planet systems contain planets in mean motion resonances.

Both theoretical and numerical analyses verify the ubiquity of planet
migration due to interaction with the protoplanetary disk of gas and
dust \citep[e.g.][]{1997Icar..126..261W, 2000MNRAS.318...18N}.
The clearing of disk material between two planets both
capable of opening a gap in the disk leads to differential migration
of the two planets as at least the outer planet is forced in by the
material remaining outside its orbit.
The convergence of the orbits naturally allows capture into stable
resonant orbital configurations.
Hydrodynamical simulations of two embedded planets capturing each
other into resonance have been performed by several groups
\citep{2000MNRAS.313L..47K,2000ApJ...540.1091B,
2001A&A...374.1092S,2003CeMDA..87...53P,2004A&A...414..735K}.
The resonance capture has also been analyzed with extensive three-body
calculations, where migration is simply imposed with either the
semi-major axis migration rate $\dot a/a$ and the eccentricity damping
rate $\dot e/e$ specified explicitly or ad hoc forces added to produce
migration and eccentricity damping 
\citep{2001A&A...374.1092S, 2002ApJ...567..596L,
2002MNRAS.333L..26N, 2004A&A...414..735K,
2004ApJ...611..517L}.

Depending on the migration rates, masses, and initial orbital
separations and eccentricities of the two planets, capture can occur
in different resonances.
For planets as massive as those in GJ 876, capture into the 2:1
resonances is robust if the initial $a_2/a_1 \la 2$ and the initial
eccentricities are small, and
it is most likely that the 2:1 resonances for GJ 876
were established by converging differential migration
\citep{2002ApJ...567..596L}.
The sequence of 2:1 resonance configurations that a system with
initially nearly circular orbits is driven through by continued
migration depends mainly on the planetary mass ratio $M_1/M_2$.
For $M_1/M_2 \approx 0.31$ as in the GJ 876 system, a system is first
captured into antisymmetric configurations with $\theta_1$ librating
about $0^\circ$ and $\Delta\varpi$ (and hence $\theta_2$) librating
about $180^\circ$.
Continued migration forces $e_1$ to larger values and $e_2$ from
increasing to decreasing until $e_2 \approx 0$ when $e_1 \approx 0.1$
Then the system converts to symmetric configurations like that of GJ
876, with both $\theta_1$ and $\Delta\varpi$ librating about
$0^\circ$ (Lee 2004; see also Fig. \ref{fig:3body1} below).
There are other types of 2:1 resonance configurations (with both
$\theta_1$ and $\Delta\varpi$ librating about $180^\circ$ or
asymmetric librations of $\theta_1$ and $\Delta\varpi$ about values
far from either $0^\circ$ or $180^\circ$) for $M_1/M_2 \approx 0.31$,
but they are either unstable for planets as massive as those in GJ 876
or not reachable by convergent migration of planets with nearly
constant masses and coplanar orbits (Lee 2004).
Asymmetric libration configurations can result from convergent
migration for 2:1 resonances with $M_1/M_2 \ga 0.95$ and for 3:1
resonances for a wider range of $M_1/M_2$ 
\citep{2003ApJ...593.1124B, 2003CeMDA..87...99F,
2004A&A...414..735K, 2004ApJ...611..517L}.

While it is easy to understand the symmetric resonance configuration
of GJ 876 from convergent migration, the small eccentricities of this
system are a puzzle.
The continued migration after capture into resonance drives the pair
of planets deeper into the resonances ($2 n_2 - n_1 < 0 $ increasing,
with $n_j=\dot\lambda_j$ being the mean orbital motions).
The identical mean retrograde precessions of the $\varpi_j$ must
therefore decrease in magnitude to keep ${\dot \theta}_j$ near zero,
i.e., to maintain the resonance configuration.
This decrease is effected by increasing the orbital eccentricities
when the system is in the configuration with both $\theta_1$ and
$\theta_2$ librating about $0^\circ$.
The three-body calculations have shown that the eccentricity growth
is rapid if there is no eccentricity damping, with the average
eccentricities ($\langle e_1\rangle = 0.255$, $\langle e_2\rangle =
0.035$) of the $\sin i=0.78$ fit of \citet{2001ApJ...551L.109L}
exceeded after only a 7\% decrease in the orbital radii after capture
\citep{2002ApJ...567..596L}.
The eccentricities can be maintained at the observed low values as
migration continues if there is significant eccentricity damping of
$\dot e/e \approx 100 \dot a/a$ \citep{2002ApJ...567..596L}.
In contrast the full hydrodynamical calculations typically give
similar timescales for both migration and eccentricity damping
\citep{2004A&A...414..735K}.
The low eccentricities become even more problematic if disk-planet
interactions drive eccentricity growth.
\citet{2003ApJ...587..398O} and \citet{2003ApJ...585.1024G} have
suggested that the eccentricity-damping corotation torques can be
reduced sufficiently to trigger eccentricity growth if the
eccentricity is above a critical value, and the eccentricity of the
outer planet in \GJ\ is well above their estimates for this critical
value.

Hydrodynamical simulations of two planets interacting with their
protoplanetary disk are computationally expensive, and the number of
such simulations in the literature is relatively small
\citep{2000MNRAS.313L..47K, 2000ApJ...540.1091B,2001A&A...374.1092S,
2003CeMDA..87...53P, 2004A&A...414..735K}.
In addition, 
often
the masses used in these simulations are not
appropriate for the GJ 876 system or the simulations did not continue
for a very long time.
While there has been more extensive three-body calculations with
imposed migration and eccentricity damping 
\citep{2001A&A...374.1092S, 2002ApJ...567..596L,
2002MNRAS.333L..26N, 2004A&A...414..735K,
2004ApJ...611..517L},
they do not model disk-planet interactions self-consistently
and, in particular, the effects of the apsidal precession induced by
the disk have not been considered.

In this paper we present a more comprehensive set of numerical
calculations treating the evolution of two planets interacting with
their protoplanetary disk, with special emphasis on the resonant system
\GJ.  For this purpose we solve the full hydrodynamical equations,
including viscous heating and radiative cooling effects in some cases,
and follow the joint planet-disk evolution over several thousand
orbits of the planets.
We find that we cannot reproduce the observed low eccentricities using
these straightforward assumptions.
Taking into account mass loss from the disk, as might be the
case in the late dissipation stages of protoplanetary disks subject
to, e.g., photoevaporation, we are able to reduce the final values of
the eccentricities close to the observed values. 

In Sect.~\ref{sec:obs} we give an overview of the
observational data for \GJ. 
In Sect.~\ref{sec:hydro-model} we explain our physical and numerical
approach, and in 
Sect.~\ref{sec:model-comput} we present the results of our simulations.
This is followed in Sect. ~\ref{sec:interpretations} with three-body
calculations and analytic theory to interpret the results of
Sect.~\ref{sec:model-comput}.
We state our conclusions in Sect.~\ref{sec:conclusion}.
\begin{table}
\caption{
Orbital parameters of the two planets of the planetary system \GJ\ at
epoch JD 2449679.6316 assuming co-planarity and $i = 90^\circ$ as
given by \citet{2004astro-ph0407441}.
The adopted stellar mass is $M_* = 0.32 \Msol$.
$P$ denotes the orbital period,  $M$ the
mass of the planet, $a$ the semi-major axis, $e$ the eccentricity,
$\pomega$ the angle of periastron at epoch, and $q$ the mass ratio
$M/M_*$.
}
\label{tab:gj876}
\begin{tabular}{c|l|l|l|l|l|l} 
\hline
   &  $P$   & $M$          & $a$     &  $e$   &  $\pomega$  & $q$ \\
   &  [d]   & [$M_{Jup}$]  &  [AU]   &        &   [deg]   &  [$10^{-3}$] \\
 \hline
 Inner &  30.38 &   0.597   &  0.13   &   0.218  &  154  &  1.78  \\
 Outer &  60.93 &   1.89    &  0.21   &   0.029  &  149   & 5.65  \\
 \hline
\end{tabular}
\end{table}
\section{The System GJ~876}
\label{sec:obs}
The first planet orbiting GJ 876 was discovered in 1998 
\citep{1998ApJ...505L.147M, 1998A&A...338L..67D}.
With the discovery of the second inner planet 
\citep{2001ApJ...556..296M}, the
near 2:1 commensurability of the orbital periods ($\approx 30$ and 60
days) implied stable mean-motion resonances, which were soon
confirmed.
The combined minimum mass of the planets is about 0.0074 of the
stellar mass ($M_* = 0.32 M_\odot$).
The large relative planetary masses and short orbital periods meant
that a dynamical fit to the radial velocity data that accounted for
the mutual gravitational interaction of the planets was required.
This was accomplished by \citet{2001ApJ...551L.109L} and 
\citet{2001ApJ...558..392R},
who used a Levenberg-Marquardt minimization scheme
driving an $N$-body integrator to find best-fit initial orbital
elements.
Recently new data obtained with the Keck Telescope have also been
included in the analysis of the system \citep{2004astro-ph0407441}, and the
latest best-fit orbital parameters are shown in Table 1 for an assumed
coplanar orbital inclination of $i = 90^\circ$.

\begin{figure}[ht]
\begin{center}
\resizebox{0.98\linewidth}{!}{%
\includegraphics{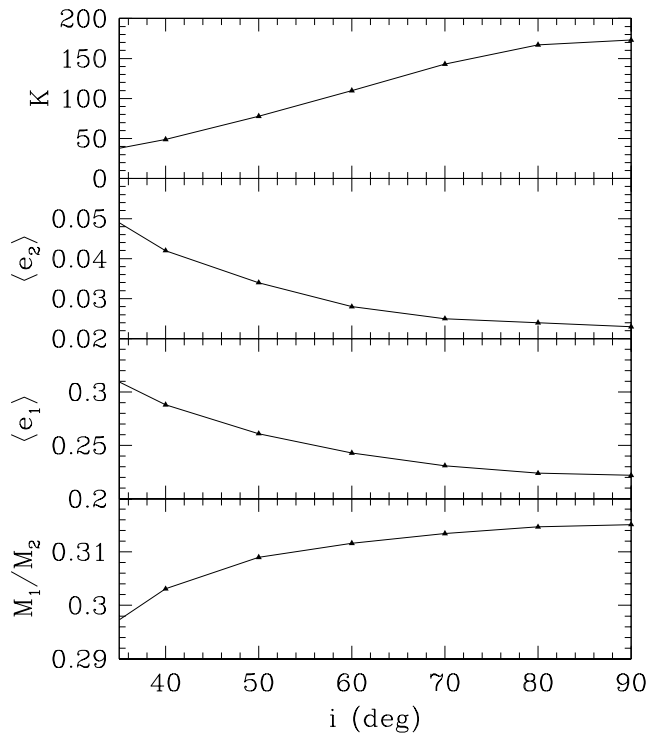}}
\end{center}
  \caption{
  Planetary mass ratio, $M_1/M_2$, average orbital eccentricities,
  $\langle e_1 \rangle$ and $\langle e_2 \rangle$, and the ratio of
  eccentricity damping rate to migration rate, $K$, for the GJ 876
  best-fit solutions of \citet{2004astro-ph0407441} with coplanar
  inclination $i > 35^\circ$.
  The value of $K = |{\dot e}_2/e_2|/|{\dot a}_2/a_2|$ is for the
  equilibrium eccentricities to match $\langle e_j \rangle$ if there
  is inward migration and damping of the outer planet only.
    }
   \label{fig:gj876}
\end{figure}

Although a dynamical fit can in principle yield the orbital
inclinations and masses of the planets, the $\chi^2$ value of the fit
for GJ 876 is relatively insensitive to the inclination $i$ of the
assumed coplanar orbits for $35^\circ \la i \le 90^\circ$ and rises
rapidly only for $i \la 35^\circ$ 
\citep{2004astro-ph0407441}.
In Fig. \ref{fig:gj876} we show the planetary mass ratio $M_1/M_2$
and the average orbital eccentricities $\langle e_1 \rangle$ and
$\langle e_2 \rangle$ for the best-fits found by 
\citet{2004astro-ph0407441} for
$i > 35^\circ$.
The deviation from Keplerian motion that is most strongly constrained
by the available observations of GJ 876 is the resonance induced
{\it retrograde} apsidal precession of the orbits at an average rate of
${\dot \varpi} = -41^\circ\yr^{-1}$, and indeed this precession
has now been observed for more than one full period.
As $i$ decreases, the individual planetary mass $M_j$ increases
roughly as $1/\sin i$ (and $M_1/M_2$ is nearly constant and $\approx
0.31$). The increase in planetary mass acts to increase ${\dot \varpi}$.
This increase can be offset by an increase in $\langle e_j \rangle$,
which acts to decrease ${\dot \varpi}$, keeping $\dot\varpi$ near the
observed value without any significant change
in $\chi^2$ until $i \la 35^\circ$.
To allow for the mass uncertainties, we shall consider systems with
different planetary masses in our numerical models.

The most striking feature of GJ 876 is the exact 2:1 orbital resonance
of the planets. The angles $\theta_1$, $\theta_2$, and
$\Delta\varpi$ describe the 
state of the system for coplanar orbits, and all three angles librate
about $0^\circ$ in the GJ 876 system.
The libration amplitudes are $|\theta_1|_{\rm max} = 7^\circ$,
$|\theta_2|_{\rm max} = 34^\circ$, and $|\Delta\varpi|_{\rm max} =
34^\circ$ for the $i = 90^\circ$ fit shown in Table 1, and the
smallest libration amplitudes in the range $35^\circ \le i \le
90^\circ$ are $|\theta_1|_{\rm max} \approx 5^\circ$ at $i \approx
50^\circ$ and $|\theta_2|_{\rm max} \approx 16^\circ$ and
$|\Delta\varpi|_{\rm max} \approx 13^\circ$ at $i \approx 35^\circ$
\citep{2004astro-ph0407441}.
The small libration amplitudes indicate that the system is deep in the
resonance, which occurs naturally in the differential migration scheme
for forming the resonant structure if the planets approach the
resonance slowly with initially small eccentricities 
\citep{2002ApJ...567..596L}.

\citet{2002ApJ...567..596L} have shown using three-body integrations with
imposed migration and eccentricity damping that the eccentricities
reach equilibrium values that remain constant for arbitrarily long
migration within the resonances if ${\dot e}_j/e_j = -K |{\dot
a}_j/a_j|$.
For inward migration and damping of the outer planet only, $K \approx
100$ is required to produce equilibrium eccentricities that match the
eccentricities of the $\sin i = 0.78$ (or $i = 51^\circ$) fit of
\citet{2001ApJ...551L.109L}.
We have repeated the calculations in 
\citet{2002ApJ...567..596L} for the updated fits
in \citet{2004astro-ph0407441}, and the value of $K$ as a function of $i$
is shown in Fig. \ref{fig:gj876}.
For $i = 51^\circ$, $K \approx 80$, which is close to the value
inferred by \citet{2002ApJ...567..596L}.
\citet{2002ApJ...581L.115B} have claimed from HST astrometric measurements
that the inclination is very close to $90^\circ$, which would require
a significantly larger amount of eccentricity damping ($K \approx
170$).
The increase in $\langle e_j \rangle$ with decreasing $i$ leads to a
decrease in $K$, but $K$ must still be greater than about 40 if $i \ga
35^\circ$.

%
%
%
\section{The Hydrodynamical Model}
\label{sec:hydro-model}
Even though the physical setup is different from existing
simulations, the models presented here are calculated 
roughly in the same manner as those described previously in
\citet{1998A&A...338L..37K, 1999MNRAS.303..696K}
for single planets and in \citet{2000MNRAS.313L..47K}
for multiple planets. 
The reader is referred to those papers
for details on the computational aspects of this type of simulations.
Other similar models, following explicitly the motion of single and
multiple planets in disks, have been presented by
\citet{2000MNRAS.318...18N}, \citet{2000ApJ...540.1091B},
\citet{2001A&A...374.1092S},
\citet{2003CeMDA..87...53P},
and more recently for resonant configurations
by \citet{2004A&A...414..735K}.
 
We use cylindrical coordinates ($r, \varphi, z$) and consider a
vertically averaged, infinitesimally thin disk located at $z=0$,
where the origin of the coordinate system is at the position of the
star.  The basic hydrodynamic equations (mass and momentum
conservation) describing the time evolution of such a two-dimensional
disk with embedded planets have been stated frequently and are not
repeated here \citep[see][]{1999MNRAS.303..696K}.  Additional
information on the treatment of embedded planets is given in
\citet{2004A&A...414..735K}.
\subsection{Energy Equation}
The majority of models presented here use a fixed temperature distribution which
follows from  the assumption of a constant ratio of vertical height $H(r)$ to
radial distance $r$ from the star. Here we assume $H(r)/r = {\rm const.} = 0.05$,
from which $T(r) \propto r^{-1}$ follows. In this case the pressure $p$
is given by $p = \Sigma c\subscr{s}^2$, where $\Sigma$ is the surface density
and $c\subscr{s}$ the isothermal sound speed. 
In the situation of a given $H/r$-ratio there is no need for solving an extra
energy equation, and we refer to those models as {\it isothermal}
(even though the radial temperature is varying).

We also present {\it radiative}
models with an improved thermodynamic treatment using
the thermal energy equation
\begin {equation}
\label{eq:energy}
 \doverdt{\Sigma c\subscr{v} T} + \nabla \cdot (\Sigma c\subscr{v} T {\bf u} )
                =  - p  \nabla \cdot {\bf u}  +  D  - Q
\end{equation}
Here, ${\bf u} = (u_r, u_{\varphi})$ is the two-dimensional velocity,
$T$ the (midplane) temperature of the disk,
$c\subscr{v}$ the ratio of specific heats,
$D$ the dissipation function, and $Q$ is the local radiative 
cooling.
The form of the dissipation function $D$ in cylindrical coordinates
can be found in \citet{1984frh..book.....M}, and to calculate the radiative
losses (from the two sides of the disk) we follow 
\citet{2003ApJ...599..548D} and \citet{2004A&A...423..559G}
and write
\begin{equation}
    Q  =  2  \sigma\subscr{R}  T\subscr{eff}^4
\end{equation}
where $\sigma\subscr{R}$ is the Stefan-Boltzmann constant and $T\subscr{eff}$ is an
estimate for the effective temperature \citep{1990ApJ...351..632H}
\begin{equation}
    T^4\subscr{eff} \, \tau\subscr{eff} = T^4   \qquad  \mbox{with}
 \qquad   \tau\subscr{eff}  
   = \frac{3}{8} \tau  + \frac{\sqrt{3}}{4} + \frac{1}{4 \tau}
\end{equation}
For a two-dimensional disk we approximate the mean vertical optical
depth by
\begin{equation}
    \tau  = \frac{1}{2}  \kappa  \Sigma
\end{equation}
where for the Rosseland mean opacity $\kappa$ the analytical formulae by
\citet{1985prpl.conf..981L} are adopted.

In addition to the simplified treatment of radiation as expressed in (\ref{eq:energy})
we also have considered models including radiative diffusion in the $(r, \varphi)$-plane.
In this case a term
\begin{equation}
\label{eq:raddif}
   - 2 H \nabla \cdot \vec{F}     \qquad  \mbox{with}
 \qquad   \vec{F}  =  -  \frac{\lambda c \, 4 a T^3}{\rho \kappa} \, \nabla T 
\end{equation}
has been added to the rhs of Eq.~(\ref{eq:energy}).
Here $\vec{F}$ denotes the radiative flux in the $(r, \varphi)$-plane, 
$c$ is the speed of light, $a$ the radiation constant, $\rho = \Sigma / (2 H)$ the
midplane density, and $\lambda$ the flux-limiter \citep[see][]{1989A&A...208...98K}.
 
We work in a rotating reference system, rotating approximately with
the initial period of the outer planet.  As the coordinate system is
accelerated and rotating, we take care to include the
indirect terms.
\subsection{Initial Setup}
The two-dimensional ($r - \varphi$) computational domain consists of a
complete ring of the protoplanetary disk centered on the star.  By
previous simulations of two embedded planets interacting with the
protoplanetary disk it has been shown that the inner part of the disk
(inside of the {\it outer} planet) clears rapidly. The final
configuration is such that two planets orbit the star inside a cavity
\citep{2004A&A...414..735K}.  In this state only the outer planet is
still in touch with the disk, see Fig.~1 in
\citet{2004A&A...414..735K}.  Taking account of this fact, we simplify
the setup and choose the radial extent of the computational domain
(ranging from $r\subscr{min}$ to $r\subscr{max}$) such that the inner
planet orbits entirely {\it inside} $r\subscr{min}$.  This reduces the
necessary radial range covered and leads to a significant reduction in
computational effort. A typical example of the grid structure is
displayed below in Fig.~\ref{fig:global}.
In the azimuthal direction for a complete annulus we
have $\varphi\subscr{min} =0$ to $\varphi\subscr{max} = 2 \pi$.

The initial hydrodynamic structure of the disk (density, temperature,
velocity) is axisymmetric with respect to the location of the star,
and the surface density scales as $\Sigma(r) = \Sigma_0 \, r^{-1/2}$,
with a superimposed initial gap \citep{2000MNRAS.313L..47K}.  The
initial velocity is pure Keplerian rotation ($u_r=0, u_\varphi = (G
M_*/r)^{1/2}$), and the initial temperature stratification is given by
$T(r) \propto r^{-1}$ which follows from an assumed constant
$H/r$.  For the isothermal models the initial temperature profile is left
unchanged, while for the radiative cases it is evolved according to
Eq.~(\ref{eq:energy}).

The kinematic viscosity $\nu$ is parameterized by an 
$\alpha$-description $\nu = \alpha c\subscr{s} H$,
where the isothermal sound-speed is given by
$c\subscr{s} = \left( c\subscr{v} T\right)^{1/2}$,
and $H(r)$ is either held fixed (for the isothermal models) 
or, for the radiative models, calculated from the temperature/sound-speed as 
$H(r) = c\subscr{s} \, / \, \Omega\subscr{K}(r)$,
where
\[
    \Omega\subscr{K} =  \left( \frac{G M_*}{r^3} \right)^{1/2} 
\]
is the Keplerian angular velocity of the disk.
\subsection{Boundary conditions}
In an effort to  ensure a  uniform environment for all models and
minimize disturbances (wave reflections) from the outer
boundary, we impose at $r\subscr{max}$ damping boundary
conditions where the density and both velocity components
are relaxed towards their initial state as
\begin{equation}
   \frac{d X}{d t}  =  - \frac{ X  - X(t=0)}{\tau\subscr{damp}}  \, R(r)^2  
\end{equation}
where $X \in \{\Sigma, u_r, u_\varphi\}$,
$\tau\subscr{damp} = 1/\Omega\subscr{K}(r\subscr{max})$ and $R(r)$
is a dimensionless linear ramp-function rising from 0 to 1 within
$[ r\subscr{damp}, r\subscr{max}]$.
This damping setup is defined in more detail in an international comparison test
project\footnote{\small \tt http://www.astro.su.se/$\tilde{~}$pawel/planets/test.hydro.html}.
As the initial radial velocity is vanishing, this damping routine
ensures that no mass flows through the outer boundary at 
$r\subscr{max}$. However, 
in some models described below, matter is allowed to leave the outer
boundary.

At the inner radial boundary $r\subscr{min}$ outflow conditions are
applied; matter may flow out, but none is allowed to enter. This
procedure mimics the accretion process onto the star. The density gradient
is set to zero at $r\subscr{min}$, while the
angular velocity there is fixed to be Keplerian.
In the azimuthal direction, periodic boundary conditions for all
variables are imposed.

These specified boundary conditions allow for a well defined
quasi-stationary state if the planets are not allowed to respond to
the disk.
\begin{figure}[ht]
\begin{center}
\resizebox{0.98\linewidth}{!}{%
\includegraphics{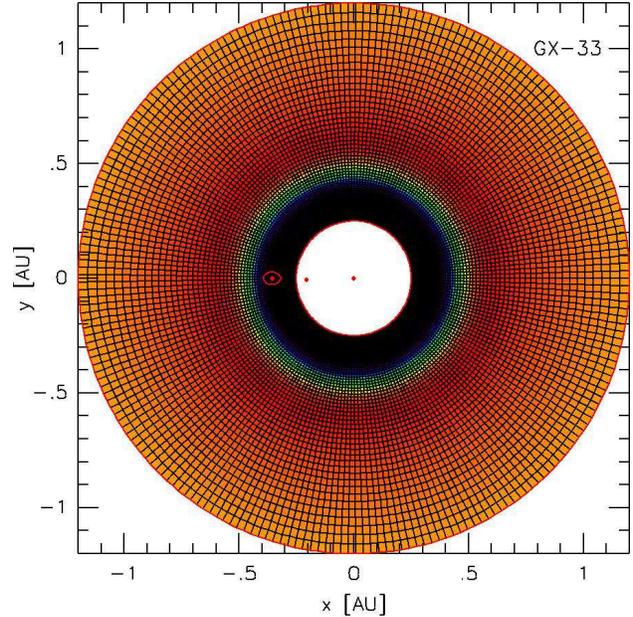}}
\end{center}
  \caption{
  The global grid structure with gray-shaded initial surface density
  superimposed.   Every second grid points is shown. 
  The dots denote the location of the star and the two planets
  and the oval line refers to the Roche lobe of the outer planet.
    }
   \label{fig:global}
\end{figure}
\subsection{Model parameters}
The computational grid in all models has the same radial extent from
$r\subscr{min}= 0.25$ to $r\subscr{max}= 1.20$AU.  It is covered by
111 $\times$ 450 ($N_r \times N_\varphi$) gridcells which are spaced
logarithmically in radius and equidistant in azimuth.  The radius
beyond which the damping procedure defined above gradually sets in is
given by $r\subscr{damp} = 1.1$.  The stellar mass is $0.32 M_\odot$
in all models.  For the planetary masses different choices have been
made as the exact parameters for \GJ\ are not known.  The majority of
models assumed an edge on system $i = 90\degx$ which leads to (see
Tab.~\ref{tab:gj876}) $m_1 = 0.597 M\subscr{Jup}$ and $m_2 = 1.89
M\subscr{Jup}$.  
Recall that the index 1 refers to the inner and the index 2 to
the outer planet.  In terms of their mass ratios
($M/M_*$) these values are equivalent to $q_1 = 1.78 \times
10^{-3}$ and $q_2 = 5.65 \, \times 10^{-3}$. In the analysis of the models we
prefer to state the planetary masses in terms of mass ratios rather
than the actual values. Most of the mass ratios used in this investigation
are close to the edge-on case.  The two planets are placed initially
at semi-major axes of $a_1 = 0.20$ AU and $a_2 = 0.35$ AU.  Both planets
have an initial eccentricity of 0.01 which is comparable to what is
found typically in numerical simulations of disk-planet interactions.
Even if the planets were started with zero initial 
eccentricity at those $a_j$,
the dynamical interaction between the planets alone
would also lead to non-zero eccentricities of the same order.

After an initial relaxation procedure, where the orbital parameters
of the planets are not changed and the disk structure is brought 
to an equilibrium, the planets are `released' 
in all cases and are allowed to migrate (change their semi-major axes)
in accordance with the gravitational disk forces/torques exerted on them.

During the evolution, material may enter from the disk into the Roche
lobe of a planet. This material is partly removed from the simulation
to model accretion onto the planet;
for the detailed procedure see \citet{1999MNRAS.303..696K}.
In some models this mass is not added to the dynamical mass
of the planets, so that mass is not strictly conserved. In other simulations
the mass is added to that of the planet, maintaining conservation of
mass.
In Table~\ref{tab:released} below it is indicated that only one of the
presented models (h8a) has a varying dynamical mass. In general it does
not change the outcome of a simulation noticably whether 
mass is accreted onto a planet or not.

For the viscosity a value of $\alpha =0.01$ is used for all models.
This is probably on the large side for protoplanetary disks, but it
allows for a rapid evolution of the system and hence a reasonable
computational effort; a larger $\alpha$ speeds up the evolution and
migration of the planet.  It has been shown earlier that the migration
speed has no influence on final magnitude of the eccentricities
\citep{2002ApJ...567..596L} which tends to justify this approach.
In addition, a larger $\alpha$-value leads to a gap that is not so
well cleared \citep{1999MNRAS.303..696K} which will tend to induce a larger
periastron advance ($\dot \pomega$) and an increased eccentricity
damping.  Both effects presumably serve to minimize the final
eccentricity of the outer planet.

The density in the system is adjusted such that in the relaxed initial
state there is approximately $2.75 \times 10^{-3}\Msol$ of material
within the computational domain (see below).
\subsection{A few remarks on numerical issues}
The numerical method used is a staggered mesh, spatially second order
finite difference method, where advection is based on the
monotonic transport algorithm \citep{1977JCoPh..23..276V}.  The code
uses operator-splitting and is semi-second order in time.  The
computational details of the code that we employ, {\tt RH2D}, have
been described in general in \citet{1989A&A...208...98K}, and
specifically for planet calculations in \citet{1999MNRAS.303..696K}.
The use of a rotating coordinate system requires special treatment of
the Coriolis terms to ensure angular momentum conservation
\citep{1998A&A...338L..37K}, an especially important point for the
long-term calculations presented here.

The viscous terms, including all necessary tensor components, are
treated explicitly.  To ensure stability in the gap region, where
there are very strong gradients in the density, an artificial bulk
viscosity has been added, with a coefficient $C\subscr{art} = 2$. For
a detailed discussion of the viscosity related issues and tests, see
\citet{1999MNRAS.303..696K}.

The energy equation Eq.~(\ref{eq:energy}) is solved explicitly applying
operator-splitting.  The heating and cooling term $D-Q$ is treated as
one sub-step in this procedure, see \citet{2004A&A...423..559G}.
The additional radiative diffusion part in the energy equation
(\ref{eq:raddif}) is solved applying an implicit method 
to avoid possible time step limitations.
To solve the resulting matrix equations we use Successive Over Relaxation
(SOR) with an adaptive relaxation parameter \citep{1989A&A...208...98K}.

A larger mass ratio $M/M_*$ induces stronger torques and
produces low densities in the gap region.
To prevent numerical instabilitites caused by too large gradients
we have found it preferable to work with a density floor, where the density
cannot fall below a specified minimum value $\Sigma\subscr{min}$. 
For our purpose we use a value of $\Sigma\subscr{min} = 10^{-6}$ in
dimensionless values, where
the typical (initial) density is of ${\cal{O}}(1)$.
\begin{figure}[ht]
\begin{center}
\resizebox{0.98\linewidth}{!}{%
\includegraphics{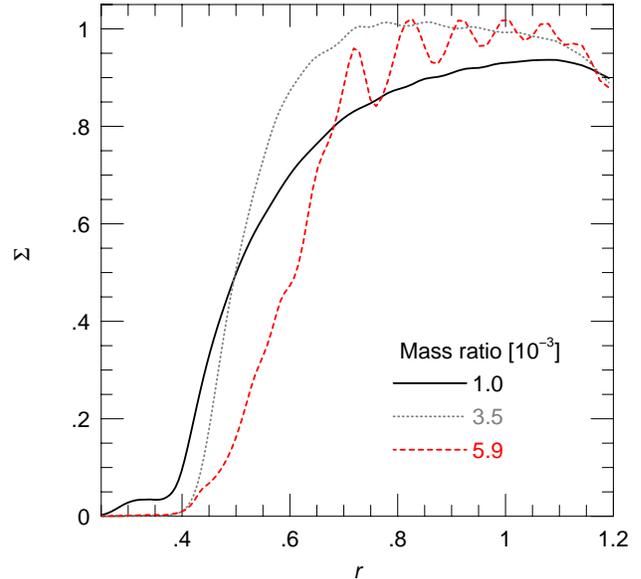}}
\end{center}
  \caption{
  The azimuthally averaged density profile for the
   relaxed configurations for three different masses of the outer planet:
  $q_2 = 1.0 \times 10^{-3}, q_2 = 3.5 \times 10^{-3}$
  and $\qtwox \times 10^{-3}$. The planet
   is located at $a_2 = 0.35$ with a fixed semi-major axis,
   and is allowed to accrete.
   In these models the mass of the inner planet has been switched off.
   The density is given in dimensionless units.
    }
   \label{fig:sigma-multi1}
\end{figure}
\subsubsection{Planetary dynamics}
The motion of the planets is integrated using a fourth order Runge-Kutta
integrator where the time step size is given by the hydrodynamical time
step. While not the most accurate integrator for full N-body calculations,
it is sufficient for our purposes. As a test we have run the pure
3-body problem 
of one star with two planets under exactly the same conditions as in the
full hydrodynamical evolution and find that the relative total energy loss
is less than $6 \times 10^{-6}$ in 1000 years 
($\approx 2.5 \times 10^{6}$ time steps),
which is equivalent to over 6,000 periods
of the inner and over around 2,700 periods of the outer planet. 

The forces of the disk are taken into account in a first order
approximation to reduce the computational effort. 
To avoid problems with under-resolved material close to the planet,
a torque cutoff radius of $r\subscr{torq}$ is applied where
material closer to the planet than $r\subscr{torq}$ does not contribute to
the force acting on the planet.
The problem of choosing the optimum value for $r\subscr{torq}$
is non-trivial. 
Using a cutoff radius prevents large unphysical variations of
the forces acting on the planet. Very high resolution
simulations \citep{2002A&A...385..647D, 2004astro-ph0411705}
show a nearly symmetric distribution
of material close to the planet indicating that those regions
do not contribute too much to the total torque.
As these regions cannot be resolved in our simulations we have to use
a torque cutoff instead.
For all isothermal models we use $r\subscr{torq} = 0.5 R\subscr{Hill}$,
where the Hill radius is given by 
\beq
    R\subscr{Hill}  =  a\subscr{2} \, \left( \frac{q_2}{3} \right)^{1/3}.
\eeq
Tests with an increased value of $0.75 R\subscr{Hill}$ gave indistinguishable
results on the migration rate and eccentricity evolution. 
For the radiative models we use a value of $r\subscr{torq} = 0.8 R\subscr{Hill}$
because the radiative cooling leads to more material around the planet.

In calculating the gravitational potential of the disk and planet
the vertical extent of the disk is taken into account,
assuming a vertically isothermal structure, i.e. Gaussian density
distribution.  
For the smoothing length of the potential we choose 
$r\subscr{pot} = 0.8 H$.  For a $1.9 \Mjup$ planet this is equivalent to
only about $0.3 R\subscr{Hill}$.
\section{Model computations}
\label{sec:model-comput}
Constructing our models for the formation of \GJ\ consists of two basic steps:\\
I. Construct equilibrium models where the planets have fixed
orbital elements, and let the disk evolve into a quasi-stationary state.\\
II. Follow the subsequent evolution by `releasing' the planets,
i.e. by taking into account the disk forces acting on the planet.
\\
This two-fold procedure is necessary to avoid evolutions which
might be dominated by a transient adjustment of the disk,
as the disk-planet equilibrium state is not known a priori.
The disadvantage is that the equilibration phase may take 
many hundreds of orbits.
Let us consider these two steps in turn.
\begin{figure}[ht]
\begin{center}
\resizebox{0.98\linewidth}{!}{%
\includegraphics{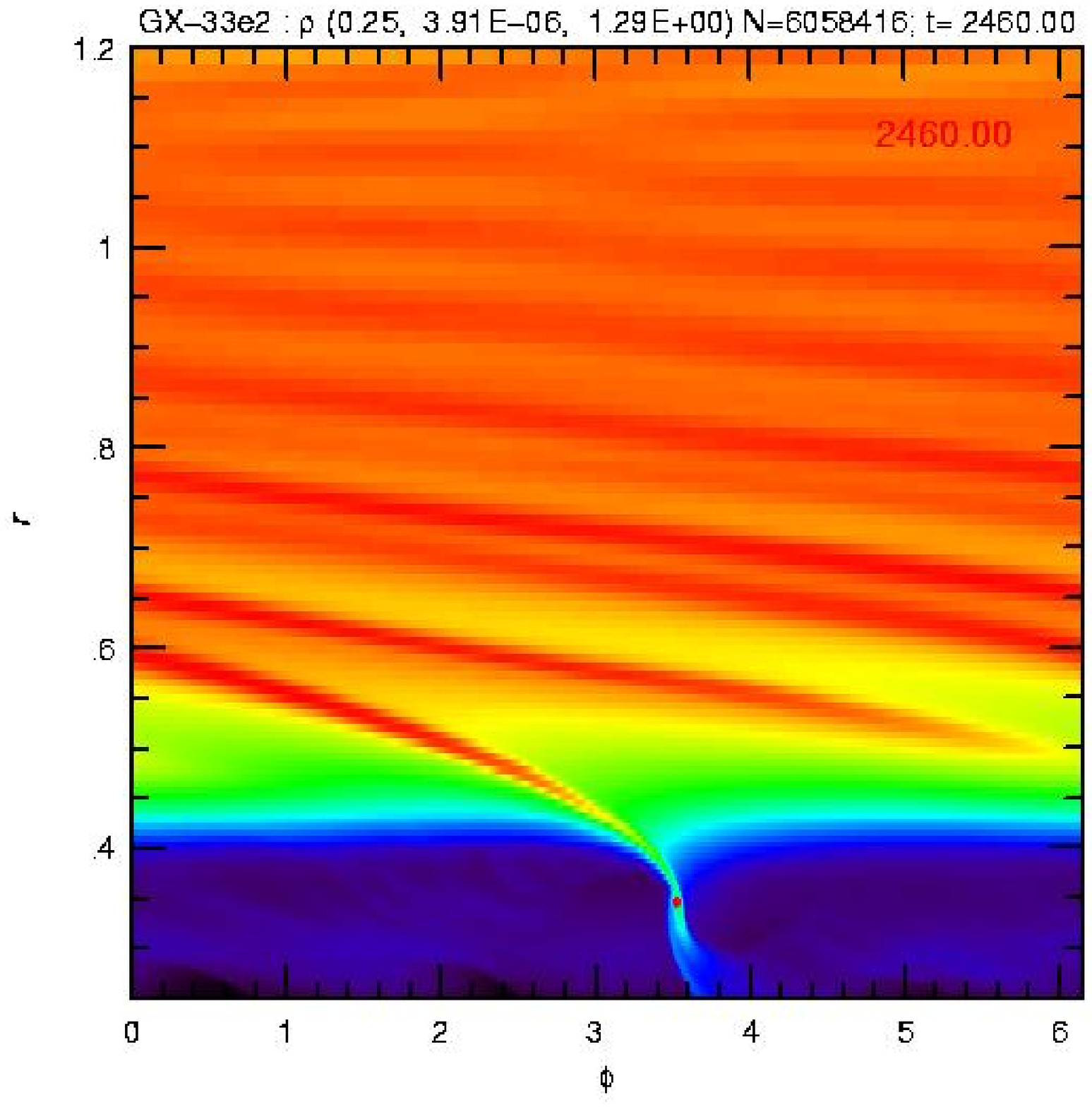}}
\resizebox{0.98\linewidth}{!}{%
\includegraphics{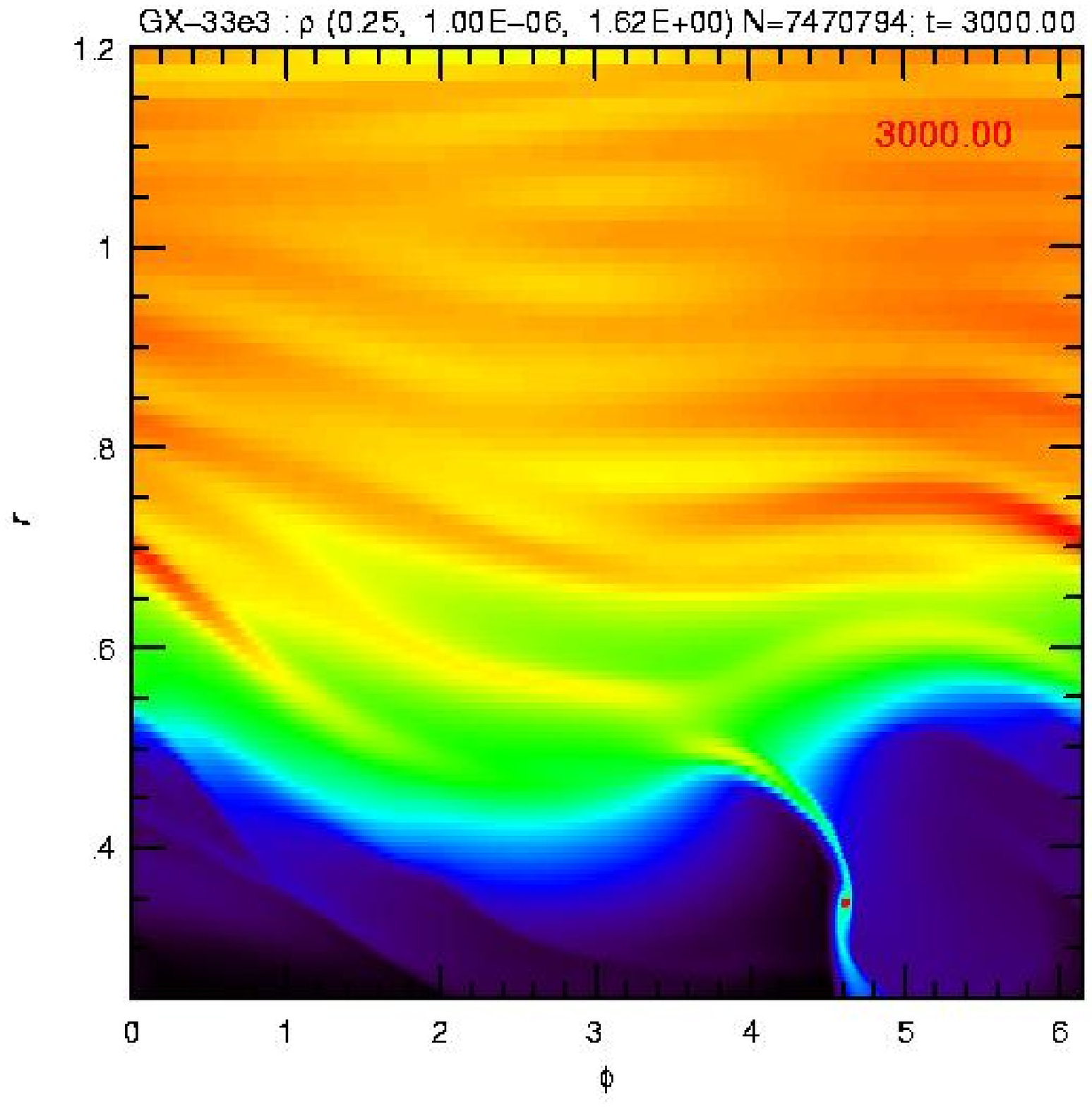}}
\end{center}
  \caption{
  Gray scale plots of the surface density $\Sigma$ for the relaxed state
   with no inner planet for two different masses:
 {\bf top}) $q_2 = 3.5 \times 10^{-3}$
   and {\bf bottom}) $q_2 = \qtwox \times 10^{-3}$.
  Due to the higher planetary mass much stronger wave-like disturbances are
  created in the density, and the disk becomes eccentric with
  very small pattern speed in the inertial frame.
   \label{fig:sigma2d}}
\end{figure}

\subsection{Relaxation towards equilibrium}
\label{subsec:equi}
\subsubsection{Switched off inner planet}
In the first part of step I the mass of the inner planet is switched
off (reduced by a factor $10^{-6}$) and the outer planet has fixed
orbital elements ($a_2 =0.35, e_2 =0.01$).  
We construct equilibrium models for different planet masses $q_2$ because 
firstly, the observations do not yield definite masses of the planets
due to the poorly constrained inclination, and secondly we want to
understand the planet-disk system in more general terms.
If the surface density and all other variables are fixed at the outer
boundary there corresponds (for a given viscosity and vertical height of the
disk) one particular equilibrium state to each mass ratio.

In Fig.~\ref{fig:sigma-multi1} the
azimuthally averaged $\Sigma(r)$ profile is shown for the relaxed
equilibrium disk states for three different masses $q_2$ of the outer planet
starting from $q_2 = 1.0 \times 10^{-3}$ (black solid line), over
$3.5 \times 10^{-3}$ (gray dotted line) upto
$\qtwox \times 10^{-3}$ (dashed line), with the last value close to the
minimum mass of the outer planet of the GJ 876 system. 
One notices that for the two smaller planetary masses the outer disk is
much more ``quiescent'' than for the large mass.
Larger planetary masses apparently lead to a restructuring of the disk.
The wave-like disturbances in the disk (seen as the
oscillatory behavior in the curves) are significantly stronger for
$q_2 \gsim 5.2 \times 10^{-3}$.

The existence of the two different equilibrium states of the disk is further
illustrated in Fig.~\ref{fig:sigma2d} where we display gray scale
plots of the surface density $\Sigma$ for the relaxed state with no
inner planet for two different masses ($q_2 = 3.5$ and $\qtwox \, \times
10^{-3}$).  In both cases the small but non-vanishing eccentricity of
the outer planet ($e_2 = 0.01$) leads to wave-like disturbances in the
disk oscillating with the planetary period.  But while for the lower
mass case ($q_2 = 3.5 \times 10^{-3}$) the disk structure remains quite
regular, the second high mass case ($q_2 = \qtwox \times 10^{-3}$) shows a
strongly disturbed disk which has gained a significant eccentricity
of about 0.25 near the gap edge.  The transition
from the non-eccentric state to the eccentric state which is here a
function only of the planetary mass may depend also on the viscosity
and temperature on the disk which we have held fixed.  To check if
this effect is induced by the non-zero eccentricity of the planet we
have run a model with a vanishing eccentricity ($e_2 = 0$) and found
the same behavior.  
We find that the eccentric disk is nearly stationary in the inertial
frame.
The mass flow onto the planet and through the gap is
higher than the flow for the non-eccentric disk that prevailed
for the lower mass planet.
While it would be interesting, a further
exploration of these features is beyond the scope of this paper,
and we leave it to a subsequent study. 
This disk eccentricity may be closely related to the disk eccentricity
induced by resonant interaction of the outer 3:1 eccentric Lindblad
resonance of planet 2 with the disk, as discussed by
\citet{2001A&A...366..263P}, and to the problem of the inner disk in
cataclysmic variables by \citet{1991ApJ...381..259L}.
This violent interaction of a massive planet with the disk explains also
some of the outlined numerical difficulties and requirements
(such as density floor and artificial viscosity).
\begin{figure*}[ht]
\begin{center}
\resizebox{0.98\linewidth}{!}{%
\includegraphics{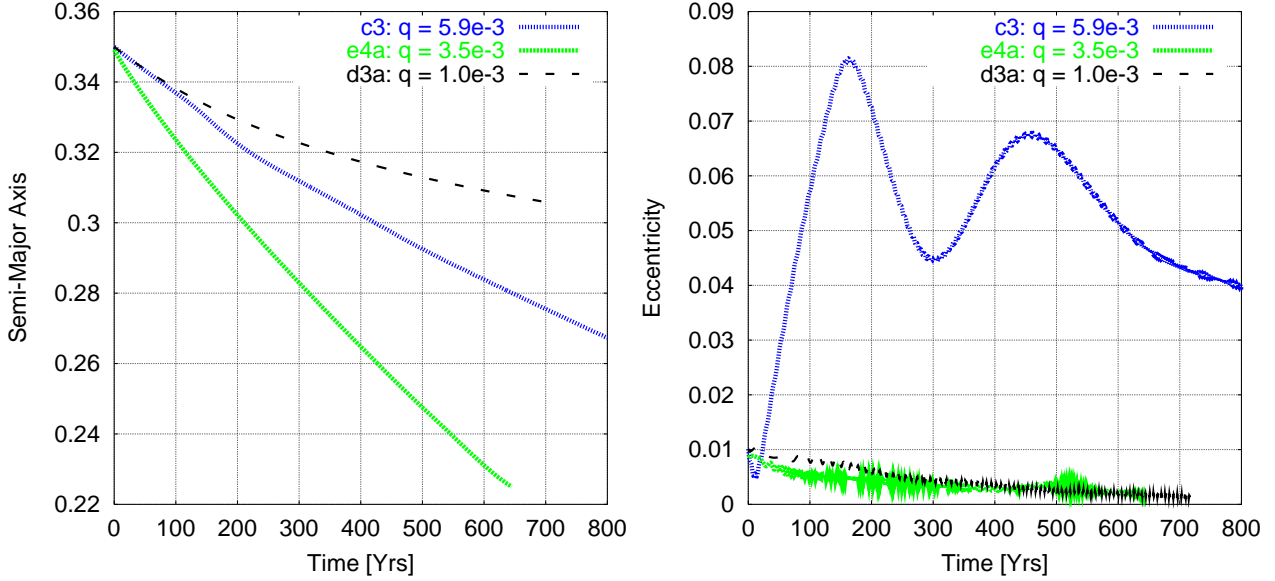}}
\end{center}
  \caption{
  The time evolution of the semi-major axis and eccentricity for three
  models with different masses of the outer planet, and zero mass inner planet.
    }
   \label{fig:aec4}
\end{figure*}
\begin{figure}[ht]
\begin{center}
\resizebox{0.98\linewidth}{!}{%
\includegraphics{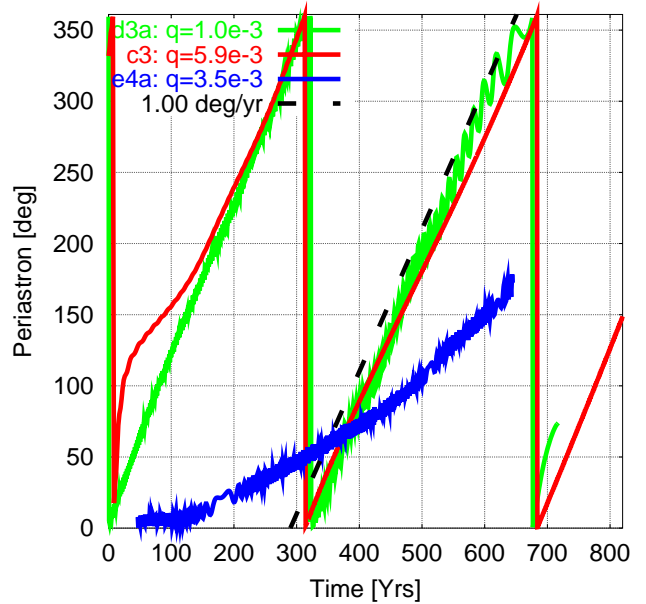}}
\end{center}
  \caption{
  The time evolution of the periastron for three
  models with different masses of the outer planet.
  The thick dashed line is a fit corresponding to
  a 1 deg/yr shift. 
    }
   \label{fig:oc4}
\end{figure}
\subsubsection{Evolution of the outer planet alone}
\label{subsec:outeralone}
To explore the effect the disk alone has on the evolution
of the orbital elements of the single (outer) planet,
in particular the influence on the precession rate, we present some
calculations where the gravitational back-reaction of the disk is
taken into account, while the inner planet is not present.
For this purpose we chose the density unit such that $\Sigma =1$ (e.g. in
Fig.~\ref{fig:sigma-multi1}) refers to a surface density of
approximately 7500 g cm$^{-2}$.
This value refers to a total mass of $2.75 \Mjup$ within the
computational domain for the $q_2 = 0.0059$ case.

\begin{figure}[ht]
\begin{center}
\resizebox{0.98\linewidth}{!}{%
\includegraphics{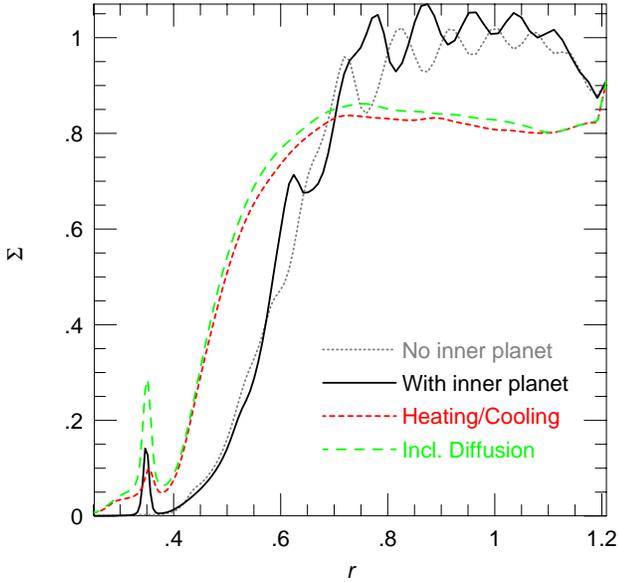}}
\end{center}
  \caption{
  The relaxed azimuthally averaged density profile for two isothermal models with
  and without considering the inner planet (solid, dotted lines),
  and 
two radiative models including only heating and cooling (short-dashed),
  and additionally radiative diffusion (long-dashed).
  The mass of the inner planet is  $q_1 = 1.75 \times 10^{-3}$ and that of the
  outer $q_2 = \qtwox \times 10^{-3}$.
 The gray short-dashed line is identical to the corresponding line
  in Fig.~\ref{fig:sigma-multi1}
  (labeled 5.9).
    }
   \label{fig:full-sig}
\end{figure}

In Figs.~\ref{fig:aec4} and \ref{fig:oc4} the time evolution of the
orbital elements of the outer planet for three models with different
planet masses are displayed. All models have been relaxed before
releasing the planet. For direct comparison the physical disk mass in all
three cases has been scaled to be identical
($\approx 2.75 \times 10^{-3} \Msol$). 
With respect to the reference model (c3) with $q_2 = 0.0059$, 
this implies for models (d3a) and (e4a) a density reduction factor of
0.98 and 0.9, respectively (cp. Fig.~\ref{fig:sigma-multi1}).
The migration rate of the planets depends primarily on
the amount of mass near the 2:1 Lindblad resonance.
We observe that the lowest ($q_2 = 0.001$) and highest ($q_2 =
0.0059$) mass planet have very similar initial migration rates, while the
intermediate mass planet has a faster migration rate.
The intermediate mass ($q_2 = 3.5 \times 10^{-3}$) has the largest
rate because there is more mass very close to the location of the
2:1 resonance, which lies here at $r = 0.56$
(note the steep density gradient in Fig. \ref{fig:sigma-multi1}).
From the average density profile one might have suspected a smaller
migration rate for the massive planet. However, due to the eccentric disk
(Fig.~\ref{fig:sigma2d}), material gets close to the planet every orbit,
which increases the migration rate slightly.
Thus, the non-monotonic variation of migration speed with planet mass is
related to the different equilibrium disk states, circular and eccentric.
The right panel of Fig.~\ref{fig:aec4} shows a temporary
increase of the eccentricity followed by a decline for the
larger mass planet case, while the smaller masses both show declining
eccentricities at nearly the same rates indicating the standard
eccentricity damping in planet-disk interaction.  This different behavior of
the eccentricity evolution is definitely caused by the eccentric disk state
for the higher planet mass case.  It remains to be studied if this
effect may have some relation to the observed large eccentricities of the
extrasolar planets.

In Fig.~\ref{fig:oc4} the evolution of the periastron is displayed
for the three cases.
For comparison the superimposed black dashed-line refers to a
periastron advance of exactly $1^\circ/$yr.
This positive $\dot{\pomega}\subscr{disk} \approx 1^\circ/$yr (for
the small and high mass planet) is solely due to the
interaction of the disk with outer planet alone.
The amount of planetary precession induced by the presence of the
disk is determined primarily by the material being very close to the planet.
Here, the behaviour of the low and high mass models are comparable again, while
for the intermediate mass, there is less material very close to the
planet and the precession rate is smallest.

\begin{figure}[ht]
\begin{center}
\resizebox{0.98\linewidth}{!}{%
\includegraphics{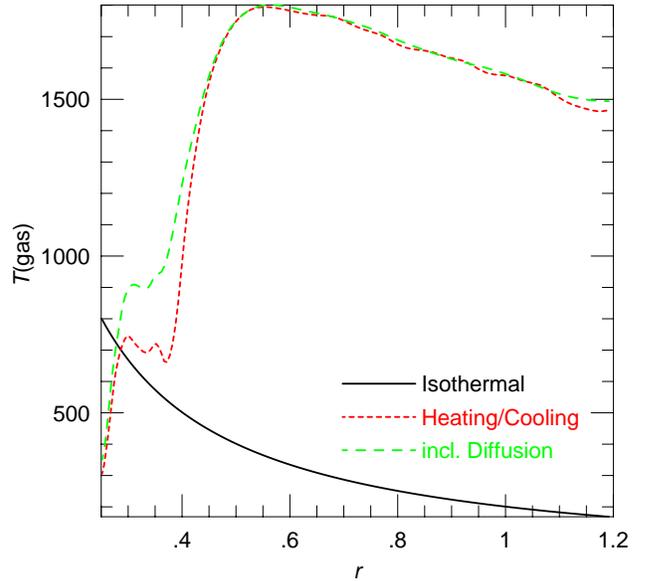}}
\end{center} 
  \caption{
  The relaxed azimuthally averaged temperature profile for the isothermal
  (solid line) and 
two radiative models including only heating and cooling (short-dashed),
  and additionally radiative diffusion (long-dashed).
  The mass of the inner planet is  $q_1 = 1.75 \times 10^{-3}$ and that of the
  outer $q_2 = \qtwox \times 10^{-3}$.
    }
   \label{fig:full-t}
\end{figure}
\subsubsection{Switched on inner planet}
In the second step of the relaxation process the inner planet is
included, using $q_1 = 1.75 \times 10^{-3}$.  Additionally, we use the
full thermodynamics for some models.  The effect that these additions
have on the density and temperature distribution is displayed in
Figs.~\ref{fig:full-sig} and \ref{fig:full-t}.  The surface density is
slightly increased upon including the inner planet because the
additional torques tend to push the matter a bit further away from the
central star.  The presence of the outer planet is now seen clearly in the
surface density distribution because for these models the planet is
not allowed to accrete material from its Roche-lobe. The subsequent
accumulation of gas in the Roche-lobe appears as a spike near $r=0.35$
in Fig.~\ref{fig:full-sig}.  
The inclusion of radiative diffusion (Eq.~\ref{eq:raddif}) in the plane of the disk
reduces the temperature in the region of the planet
and leads to a larger mass accumulation in the Roche lobe of the planet than in the model
which includes only heating and cooling (Eq.~\ref{eq:energy}). The density distribution
in the disk and gap region are not influenced too much by
including radiative diffusion. 

\begin{figure}[ht]
\begin{center}
\resizebox{0.98\linewidth}{!}{%
\includegraphics{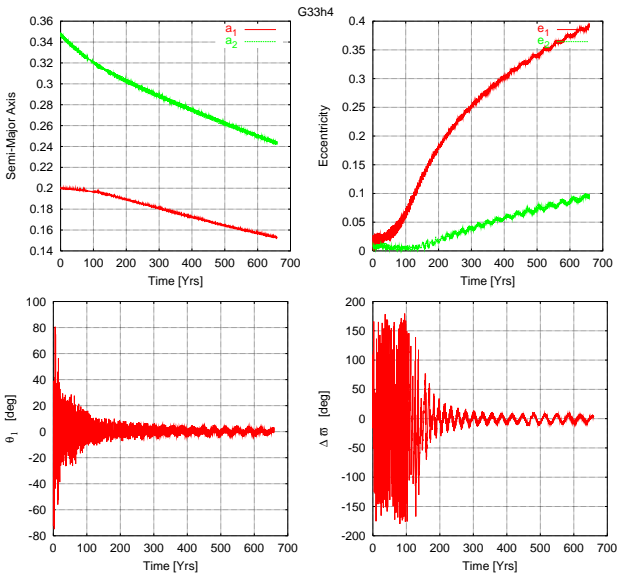}}
\end{center}
  \caption{
 The time evolution of the orbital elements ($a, e, \theta_1, \Delta \pomega$)
 for an isothermal model (h4)
 with $q_1 = 1.75\times 10^{-3}$ and $q_2 = 3.5\times 10^{-3}$ 
 (test case with lower outer planet mass).
    }
   \label{fig:g33h4}
\end{figure}
\begin{figure}[ht]
\begin{center}
\resizebox{0.98\linewidth}{!}{%
\includegraphics{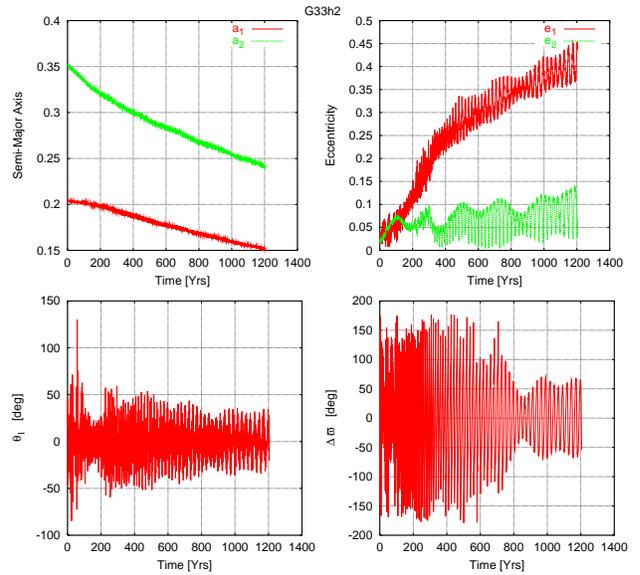}}
\end{center}
  \caption{
  The time evolution of the orbital elements ($a, e, \theta_1, \Delta \pomega$) for an isothermal
  model (h2) with $q_1 = \qonex\times 10^{-3}$ and $q_2 = \qtwox\times 10^{-3}$ (as GJ~876 edge on). 
    }
   \label{fig:g33h2}
\end{figure}
\begin{figure}[ht]
\begin{center}
\resizebox{0.98\linewidth}{!}{%
\includegraphics{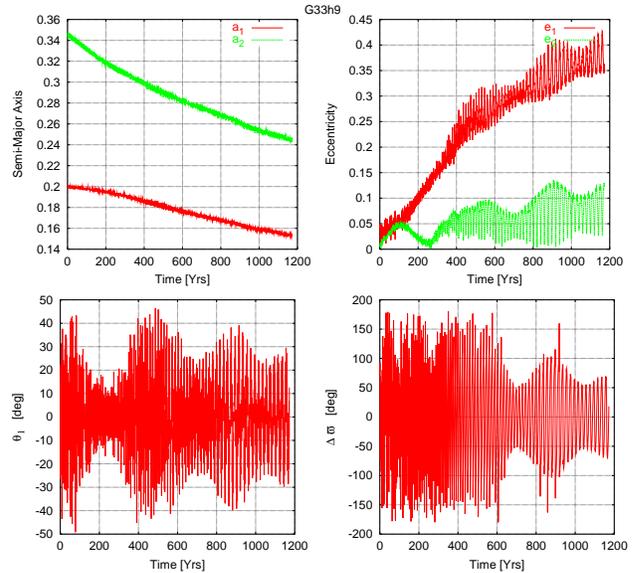}}
\end{center}
  \caption{
  The time evolution of the orbital elements ($a, e, \theta_1, \Delta \pomega$) for an isothermal
  model (h9) with $q_1 = 2.10\times 10^{-3}$ and $q_2 = 7.1\times 10^{-3}$ (as GJ~876 at $55\deg$ inclination). 
    }
   \label{fig:g33h9}
\end{figure}

\begin{figure}[ht]
\begin{center}
\resizebox{0.98\linewidth}{!}{%
\includegraphics{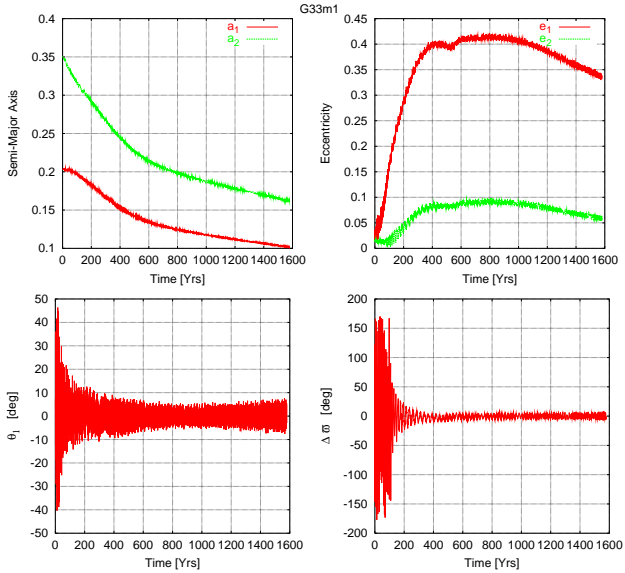}}
\end{center}
  \caption{
  The time evolution of the orbital elements ($a, e, \theta_1, \Delta \pomega$) for a 
  \underline{radiative model} including heating and cooling
 (m1) with $q_1 = \qonex\times 10^{-3}$ and $q_2 = \qtwox\times 10^{-3}$
  (as GJ~876 edge on). Note
  that at around $t=380$ the outer planet leaves the computational grid and the
  results are not reliable anymore.
    }
   \label{fig:g33m1}
\end{figure}

In Fig.~\ref{fig:full-t} the change in temperature due to the
inclusion of radiative effects is displayed.  Due to the relatively
large density of disk and the high viscosity,
the temperature in the
radiative cases rise
considerably above the isothermal case, and is given
in the disk region
by the equilibrium of heating and cooling $D=Q$.  The disk thickness
increases from $H/r = 0.05$ to about $H/r = 0.15$, for the full
radiative models. 
Including radiative diffusion in the plane of the disk (long-dashed line)
leads to higher temperature in the gap region.
The full thermodynamic models are no longer eccentric 
even for the large planetary mass.
The included radiation and larger $H/r$ leads to 
a narrower gap and additional damping
of the modes. We do not investigate at this point the detailed dependence
of the disk thermodynamics on variations of the physical viscosity.

These fully relaxed models including the inner planet
serve as the starting point for the dynamical evolution of the
planets.  The periastron advance for the outer planet due to the inner
planet having $q_1 = 0.00175$ and $a_1=0.20$  alone, with {\em no} disk
forces is found to be $\dot{\pomega}\subscr{planet}
\approx 0.68\degx/$yr. Thus, the disk and the planet generate a
shift of similar magnitude.
%
\subsection{Evolving planets}
\label{subsec:evolv}
Having constructed several equilibrium models holding the orbital
elements of the planets fixed, we now release the planets and follow
the evolution of their orbital elements.  For an overview
Table~\ref{tab:released} lists the models and their most important
parameters.

\begin{table}
\caption{
Parameters of the models used to study the evolution of
a planetary system under the action of an outer disk.
Given are a model name, the used thermodynamics (TD), 
isothermal (iso) or radiative (rad),
the type of outer boundary condition (obc)
(open or closed),  the mass-ratio of the inner planet ($q_1$),
the mass ratio or evolution of the outer planet ($q_2$), the figure
in which the model is displayed.
The displayed radiative models include heating and cooling only
as the inclusion of the diffusion does not change results significantly}.
\label{tab:released}
\begin{tabular}{c|c|c|l|l|l|l}
\hline
  Name  &  TD    &  obc        &  $q_1$    &  $q_2$    &  \\
        &       &              &    [$10^{-3}$]     &   [$10^{-3}$]  &  Fig. & \\
 \hline
   h4 & iso &   closed  &  1.75    &  3.50  &  \ref{fig:g33h4}  &   \\
   h2 & iso &   closed  &  1.75    &  \qtwox &  \ref{fig:g33h2}  &    \\
   h9 & iso &   closed  &  2.10    &  7.10  &   \ref{fig:g33h9}  &   \\
   m1 & rad &   closed  &  1.75    &  \qtwox &  \ref{fig:g33m1}  &   \\ 
   k3b & iso &   open  &  1.75    &  \qtwox  &  \ref{fig:g33k3b}  &   \\ 
   h8a & iso &   open  &  2.10    &  \qtwox $\rightarrow$  6.9  &
     \ref{fig:g33h8a}  &    \\
   m2 & rad &   open  &  1.75    &  \qtwox  &  \ref{fig:g33m2}  &   \\ 
 \hline
\end{tabular}
\end{table}

In the first instance we keep the damped and reflecting boundary
conditions at the outer boundary, i.e. we model the situation where
the disk remains in full contact with the planet during the evolution.
In Figs.~\ref{fig:g33h4} to \ref{fig:g33h9} we display the evolution
of the semi-major axes, the eccentricities, and the two resonant angles
$\theta_1$ and $\Delta \pomega$ for three models, a test case (model h4)
with a lighter outer planet ($q_1 = 0.00175, q_2 = 0.0035$),
a case (h2) resembling the edge-on ($q_1 = 0.00175, q_2 = 0.0059$) system,
and finally the $i = 55\degx$ ($q_1 = 0.0021,
q_2 = 0.0071$) case (h9), respectively. In all cases the outer planet
captures the inner one into a 2:1 resonance.
Although the orbits are initially quite far from 2:1 mean-motion
commensurability, the resonant angle $\theta_1$ is already in
libration, with amplitude as large as $\sim 90^\circ$.
However, this libration of $\theta_1$ has little dynamical consequence
and the increase in $e_1$ is slow, until the mean motions approach the
2:1 commensurability.
There is a delay in the capture of $\Delta\varpi$ (and $\theta_2$)
into resonance, and the libration amplitude of $\theta_1$ is smaller
than that of $\Delta \pomega$.
Note that in the first test-case for smaller planet mass, when the
initial disk state was not eccentric, the capture of $\Delta \varpi$
occurs faster and the libration amplitudes of both $\theta_1$ and
$\Delta\varpi$ are much smaller.

The fact that $\theta_1$ (and even $\theta_2$ and $\Delta\varpi$) can
be librating when the orbital mean-motions are far from the 2:1
commensurability is due to the $1/e_j$ dependence of the
resonance-induced retrograde precession of $\varpi_j$ at small $e_j$.
The relative timing of the capture of $\theta_1$ and $\theta_2$ into
resonance is affected by the masses and initial eccentricities of
the planets.
We shall discuss these points further in Sect. \ref{sec:interpretations}.

Once both $\theta_1$ and $\Delta\varpi$ are librating about $0^\circ$,
the eccentricities rise rapidly.
In cases h2 and h9, applicable to GJ 876, the eccentricity of the
inner planet rises above 0.4, which clearly exceeds the upper limit of
$\sim 0.31$ for \GJ\ (see Fig. \ref{fig:gj876}).

The evolution with the inclusion of 
{\bf heating and cooling} (m1) is
displayed in Fig.~\ref{fig:g33m1}. Again the system is caught in a 2:1
resonance, and the eccentricities rise to high values.
Here, due to the smoother initial state, the alignment of the
resonant angles is much stronger and the
libration amplitudes for both angles are reduced strongly.
The evolution beyond $t=380\,$yr is unreliable because both planets
are outside the computational domain by then.
The models with {\bf radiative diffusion} do not yield any different results
from models with only heating and cooling. The equilibrium
configurations as shown in Figs.~\ref{fig:full-sig} and \ref{fig:full-t} 
above are very similar already.

In summary, there exist two different types of evolutions of the
orbital elements depending on the
state of disk. For a nearly circular disk (models h4, m1) the delay 
in the capture of $\Delta \pomega$ (or equivalently $\theta_2$) allows
$e_2$ to be initially damped. After capture in this
second resonance $e_2$ rises rapidly and increases far beyond the
upper limit inferred from the observations. The libration of the
resonant angles after capture is very small. This result confirms
earlier findings of hydrodynamical evolutions 
of resonant planets \citep{2004A&A...414..735K}.
For an eccentric disk state, i.e. for larger outer planet masses, capture of
$\theta_2$ is more delayed and the libration of the two resonant
angles remains quite large.
Nevertheless, the eccentricities quickly exceed the upper limits for
GJ 876.

\begin{figure}[ht]
\begin{center}
\resizebox{0.98\linewidth}{!}{%
\includegraphics{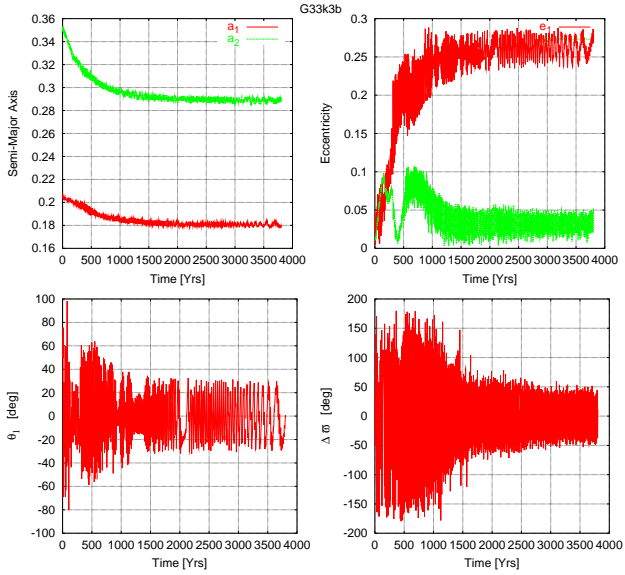}}
\end{center}
  \caption{
  The time evolution of the orbital elements ($a, e, \theta_1, \Delta
  \pomega$) for an isothermal 
  model (k3b) with $q_1 = 1.75\times 10^{-3}$ and $q_2 = \qtwox\times 10^{-3}$,
  including disk dispersal.  
    }
   \label{fig:g33k3b}
\end{figure}

\begin{figure}[ht]
\begin{center}
\resizebox{0.98\linewidth}{!}{%
\includegraphics{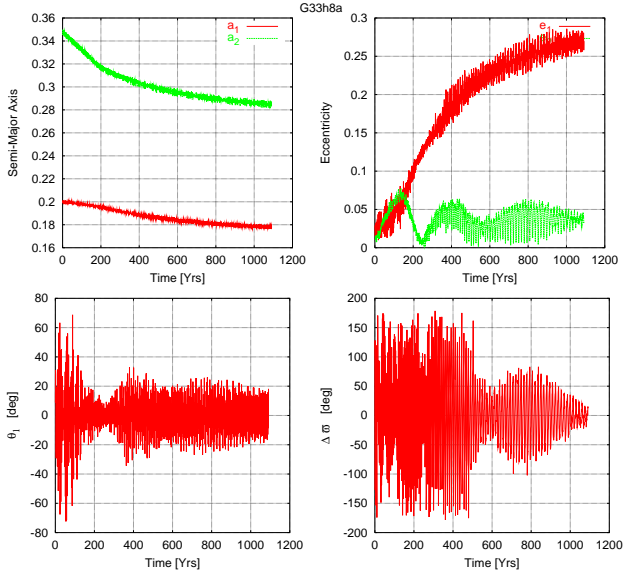}}
\end{center}
  \caption{
  The time evolution of the orbital elements ($a, e, \theta_1, \Delta
  \pomega$) for an isothermal 
  model (h8a) with $q_1 = 1.75\times 10^{-3}$ and a growing outer planet
  $q_2 = \qtwox \rightarrow 6.9\times 10^{-3}$.  The model includes disk dispersal. 
    }
   \label{fig:g33h8a}
\end{figure}

\begin{figure}[ht]
\begin{center}
\resizebox{0.98\linewidth}{!}{%
\includegraphics{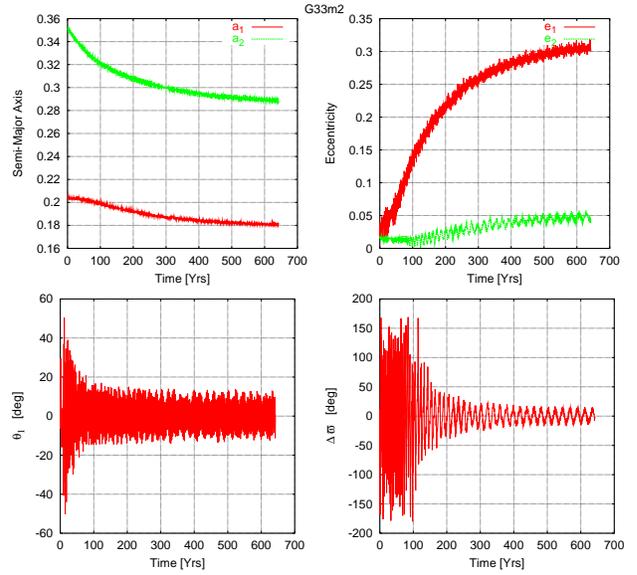}}
\end{center}
  \caption{ The time evolution of the orbital elements ($a, e,
  \theta_1, \Delta \pomega$) for a radiative model (m2) with $q_1 =
  1.75\times 10^{-3}$ and $q_2 = \qtwox\times 10^{-3}$, including disk
  dispersal.  }
   \label{fig:g33m2}
\end{figure}

\subsubsection{Modeling disk dispersal}
\label{subsubsec:disperse}
The next set of models starts with the same conditions as the previous
ones but allows for material to leave the outer boundary; this could
occur if the outer disk is photoevaporated, for example.  Depending on
the maximum outflow velocity allowed, the mass in the disk can be
reduced slowly or very efficiently. For our purposes we set the
velocity in the outer radial ghostcell (the gridcell just beyond
$R\subscr{max}$) to 1/10 of the value of the adjacent
inner gridcell. 
This results in a mass loss rate which 
is somewhat higher than physically plausible, but allows us to halt the
migration in a computationally feasible time.
For the isothermal models (k3b) and (h8) we find a mass-half-emptying
time of about 300 yrs and for the radiative model 100 yrs.

In Figs.~\ref{fig:g33k3b} and \ref{fig:g33h8a} two isothermal cases
are presented; in the first (k3b), the mass of the outer planet remains
constant, while in the second model (h8a) the planet is allowed to accrete
from its surroundings and to grow in mass. In the second case the mass
of the inner planet was chosen higher $q_1 = 0.0021$ to compensate for
the higher final mass of the outer planet.  Again, capture into the
2:1 resonance occurs, but the eccentricities do not rise to
such high values. The loss of mass from the disk turns the migration
off before the eccentricities get large. The alignment of the
resonant angles occurs on 
longer timescales than before.  Similar behavior can be seen in the
evolution of the radiative model (m2) displayed in Fig.~\ref{fig:g33m2}. Again
the alignment of the resonant angles occurs faster and
libration of the resonant angles is smaller than in the isothermal
models due to the non-eccentric disk state. 

\section{Interpretations}
\label{sec:interpretations}
In this section we use three-body integrations and analytic theory to
interpret the hydrodynamic results presented in
Sect. \ref{sec:model-comput}.
The three-body integrations were performed using the symplectic
integrator SyMBA \citep{1998AJ....116.2067D}, modified to include forced
migration and apsidal precession and to have input and output in
Jacobi coordinates \citep{2002ApJ...567..596L, 2004ApJ...611..517L}.
Although three-body integrations do not model the interactions between
the disk and the planets self-consistently, they allow different
effects to be included separately to assess their importance.
In particular, we show that the disk-induced apsidal precession, which
has not been included in previous three-body integrations, does not
cause an additional delay in the capture of $\theta_2$ and $\Delta\varpi$,
and that the additional delay in the capture of
$\Delta\varpi$ in, e.g., the isothermal models h2
(Fig. \ref{fig:g33h2}) and h9 (Fig. \ref{fig:g33h9}) is due to the
relatively large initial eccentricities induced by the eccentric disk.

\begin{figure}[ht]
\begin{center}
\resizebox{0.98\linewidth}{!}{%
\includegraphics{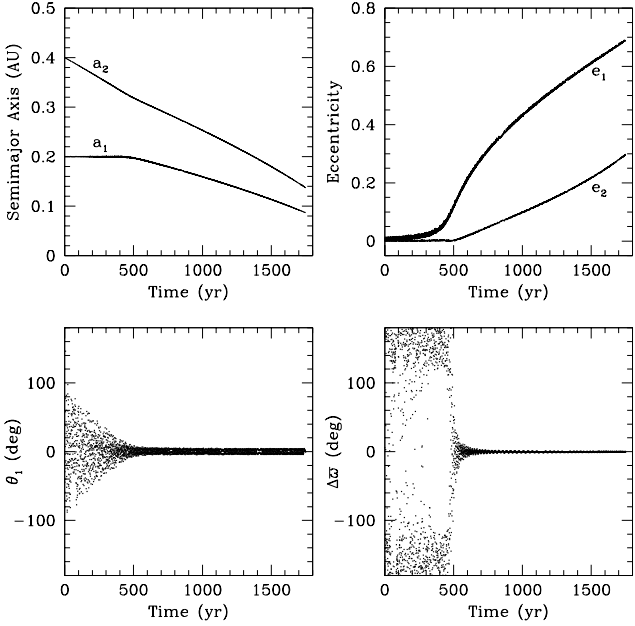}}
\end{center}
  \caption{
  Time evolution of the orbital elements ($a, e, \theta_1, \Delta
  \pomega$) for the baseline three-body model with $q_1 = 1.75 \times
  10^{-3}$ and $q_2 = 5.9 \times 10^{-3}$.
  The planets are initially on coplanar circular orbits.
  The outer planet is forced to migrate inward at the rate ${\dot
  a_2}/a_2 = -4.8 \times 10^{-4}\,(0.35 \au/a_2)^{3/2} \yr^{-1}$, and
  there is no eccentricity damping or additional apsidal precession.
  }
   \label{fig:3body1}
\end{figure}

Figure \ref{fig:3body1} shows the evolution of the semimajor axes,
eccentricities, and resonant angles $\theta_1$ and $\Delta\varpi$ for
the baseline three-body model with $q_1 = 1.75 \times 10^{-3}$ and $q_2
= 5.9 \times 10^{-3}$.
The outer planet is forced to migrate inward at the rate ${\dot
a_2}/a_2 = -4.8 \times 10^{-4}\,(0.35 \au/a_2)^{3/2} \yr^{-1}$ (which
matches the migration rate at $a_2 = 0.35 \au$ for the h2 model in
Fig. \ref{fig:g33h2}), and there is no eccentricity damping or
additional apsidal precession.
The planets are initially on coplanar circular orbits, with $a_1 =
0.2 \au$ and $a_2 = 0.4 \au$.
The circular orbits and larger $a_2$ compared to the hydrodynamical
simulations (where $a_2 = 0.35 \au$) reduce the initial eccentricity
variations.
As in the hydrodynamical simulations, the resonant angle $\theta_1$
is already librating about $0^\circ$ at the start of the evolution,
even though the planets are started even further from the 2:1
mean-motion commensurability than in the hydrodynamical simulations.
With the small initial eccentricity variations and the use of Jacobi
coordinates, it can be seen in Fig. \ref{fig:3body1} that
$\Delta\varpi$ (and $\theta_2$) are first captured into libration
about $180^\circ$ at $t \approx 270\yr$ and that $\Delta\varpi$ spends more
time around $180^\circ$ even before the capture.
The system passes smoothly over to the configuration with both
$\theta_1$ and $\Delta\varpi$ librating about $0^\circ$ when $e_1 \ga
0.1$ \citep[see also][]{2004ApJ...611..517L}.
The small libration amplitudes of the configuration with both
$\theta_1$ and $\Delta\varpi$ librating about $0^\circ$ are similar to
those found in the radiative model m1 (Fig. \ref{fig:g33m1}).

For two planets orbiting a star in coplanar orbits, the equations of
motion for the periapse longitude and eccentricity are
\begin{eqnarray}
{d\varpi_j \over dt}
&=& -{\sqrt{1 - e_j^2} \over M_j e_j \sqrt{G M_\ast a_j}}
     {\partial \Phi \over \partial e_j}
    +{\dot\varpi}_{{\rm sec},j} + {\dot \varpi}_{{\rm disk},j}
\label{eq:dwdt} \\
\frac{de_j}{dt}
&=&  \frac{\sqrt{1-e_j^2}}{M_j e_j\sqrt{G M_\ast a_j}}
     \frac{\partial \Phi}{\partial\varpi_j} \nonumber\\
& & -\frac{(1-e_j^2)-\sqrt{1-e_j^2}}{M_j e_j\sqrt{G M_\ast a_j}}
     \frac{\partial \Phi}{\partial\lambda_j}
    +{\dot e}_{{\rm disk},j} ,
\label{eq:dedt}
\end{eqnarray}
where the disturbing potential
\begin{equation}
\Phi = -{G M_1 M_2 \over a_2} \varphi (\beta, e_1, e_2, \theta_1, \theta_2) ,
\label{eq:Phi}
\end{equation}
the ratio $a_1/a_2 = \beta$, and $\varphi$ is a function of the
indicated variables if we consider only the 2:1 resonant terms and
neglect terms of order $[(M_1+M_2)/M_*]^2$ and higher 
\citep[e.g.][]{1981Icar...47....1Y, 2002ApJ...567..596L,
2003MNRAS.341..760B}.
The term ${\dot\varpi}_{{\rm disk},j}$ represents the positive apsidal
precession induced by the disk, ${\dot\varpi}_{{\rm sec},j}$
the direct secular effect from the other planet, and ${\dot e}_{{\rm
disk},j}$ the eccentricity damping induced by the disk.
The disk-induced variations are not included in Fig. \ref{fig:3body1},
and we neglect them in our discussion for the moment. 

When the eccentricities are very small, the relevant terms in $\varphi$
are $C_1(\beta) e_1 \cos\theta_1$ and $C_2(\beta) e_2 \cos\theta_2$,
and we have
\begin{eqnarray}
d\varpi_1/dt &=& \beta q_2 n_1 C_1 e_1^{-1} \cos\theta_1
                 +{\dot\varpi}_{{\rm sec},1} ,
\label{eq:dw1dt} \\
d\varpi_2/dt &=&       q_1 n_2 C_2 e_2^{-1} \cos\theta_2
                 +{\dot\varpi}_{{\rm sec},2} ,
\label{eq:dw2dt}
\end{eqnarray}
and
\begin{eqnarray}
de_1/dt &=& -\beta q_2 n_1 C_1 \sin\theta_1 ,
\label{eq:de1dt} \\
de_2/dt &=& -      q_1 n_2 C_2 \sin\theta_2 ,
\label{eq:de2dt}
\end{eqnarray}
where $n_j$ are the mean motions and $C_1(\beta) \approx -1.19$ and
$C_2(\beta) \approx +0.43$ for $\beta \approx 2^{-2/3}$ 
\citep{1981Icar...47....1Y}.
Stable simultaneous librations of both $\theta_1$ and $\theta_2$ (or
equivalently $\Delta\varpi$) require that the longitudes of the
periapses regress at the same rate on average and that the
eccentricities do not change in the absence of continued migration.
Since $C_1 < 0$ and $C_2 > 0$, the only way these requirements are
satisfied is for $\theta_1$ to librate about $0^\circ$ and $\theta_2$
and $\Delta\varpi$ to librate about $180^\circ$.
This anti-aligned configuration is what we observe with the Io-Europa
pair of Jupiter's satellites and is shown in Fig. \ref{fig:3body1}
just after the capture of $\Delta\varpi$.

The relative timing of the capture of $\theta_1$ and $\theta_2$ into
resonance needs some clarification.
Due to the $1/e_1$ dependence of the resonance-induced precession of
$\varpi_1$ at small $e_1$ (Eq. (\ref{eq:dw1dt})), a small initial
value of $e_1$ leads to an extremely mobile longitude of periapse
$\varpi_1$, such that the large mass of the outer planet can induce
sufficient retrograde motion of $\varpi_1$ to cause $\theta_1$ to
already be librating at the start of the evolution when the orbital
mean motions are far from the 2:1 commensurability.
This initial libration is evident in the three-body model shown in
Fig. \ref{fig:3body1} and in the hydrodynamical simulations shown in
Sect. \ref{subsec:evolv}.
It should be noted, however, that this libration of $\theta_1$ has
little dynamical consequence and the increase in $e_1$ (see next
paragraph) is slow, until the mean motions approach the 2:1
commensurability.
In spite of the relatively small initial value of $e_2$ in
Fig. \ref{fig:3body1} and the same $1/e$ dependence of the
resonance-induced precession (Eq. (\ref{eq:dw2dt})), $\theta_2$ and
hence $\Delta\varpi$ are not librating initially, although they do
spend more time near $180^\circ$.
This is due to the inner planet's lower mass being insufficient to
perturb the more massive outer planet's $\varpi_2$ enough to cause
$\theta_2$ to librate that far from the 2:1 commensurability of the
mean motions.

The reason the eccentricities increase with continued migration can be
understood as follows.
Continued inward migration of the outer planet while the system is
within the resonances means $a_2$ is consistently slightly smaller
that it would have been without the migration.
The increased value of $n_2$ therefore means $\lambda_2$ is slightly
larger at any instant than it would be without the migration.
The argument of the sine in Eq. (\ref{eq:de1dt}), $\theta_1 =
2\lambda_2 - \lambda_1 - \varpi_1$, is then slightly greater than
$0^\circ$ on average.
Since $C_1 < 0$, $de_1/dt > 0$ and the eccentricity must grow.
A similar argument applies to $de_2/dt$ if $\theta_2$ is also
librating about $180^\circ$ (see Eq. (\ref{eq:de2dt})).
It can be shown from the equations for the evolution of the orbital
energy and angular momentum that at least one, but not necessarily
both, of the eccentricities must be increasing at any given time for
continued migration within the resonance, if $e_2^2 < (3 + e_1^2)/4$
for the 2:1 resonance and there is no eccentricity damping \citep[see
Eq.~(A10) of][]{2002ApJ...565..608M}.

For $M_1/M_2 \approx 0.31$, the resonant interaction is no longer
dominated by the lowest order resonant terms when $e_1 \ga 0.04$, and
$e_2$ changes from increasing to decreasing with continued migration
until $e_2 \approx 0$ when $e_1 \approx 0.1$.
This change in the evolution of $e_2$ is not obvious in
Fig.~\ref{fig:3body1} because $e_2 \la 0.002$ during this phase;
see Fig. 5 of \citet{2004ApJ...611..517L}.
Then the libration center of $\theta_2$ and $\Delta\varpi$ changes
from $180^\circ$ to $0^\circ$ (with a slight phase shift due to the
continued migration), and both eccentricities continue to grow because
of the slight phase shift in the arguments of the terms involving
linear combinations of $\theta_1$ and $\theta_2$.
However, because there are now many terms in the disturbing potential
that contribute to the resonant interaction, a simple demonstration of
the co-precession of the periapses with both $\theta_1$ and $\theta_2$
librating about $0^\circ$ and of the increase in both eccentricities
with continued migration is elusive.
Continued migration causes the eccentricities to quickly exceed those
observed without a large eccentricity damping as discussed above.
Although the hydrodynamical simulations self-consistently produce
$\dot e_{\rm disk} < 0$, the magnitude of this damping is far less
than that necessary to maintain the small observed eccentricities in
the GJ 876 system during continued migration.

\begin{figure}[ht]
\begin{center}
\resizebox{0.98\linewidth}{!}{%
\includegraphics{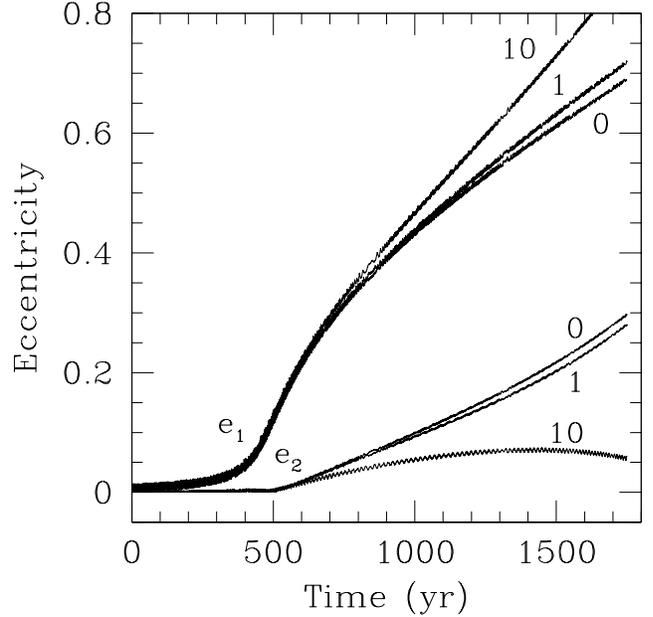}}
\end{center}
  \caption{
  Time evolution of the orbital eccentricities for the three-body
  models with additional apsidal precession ${\dot \varpi}_{{\rm
  disk},2} = p\,(0.35 \au/a_2)^{3/2} \,^\circ \yr^{-1}$ and $p = 1$
  and 10, compared to the case without additional apsidal precession
  ($p = 0$) from Fig. \ref{fig:3body1}.
  }
   \label{fig:3body2}
\end{figure}

\subsection{Disk-Induced Apsidal Precession}
It is shown in Fig. \ref{fig:oc4} that the disk induces a {\it
prograde} apsidal precession of the outer planet of about $1^\circ
\yr^{-1}$ for $q_2 = 5.9 \times 10^{-3}$.
This prograde apsidal precession is primarily due to the axisymmetric
component of the disk potential.
For a planet with orbital semimajor axis $a$ and mean motion $n$,
an outer disk with surface mass density $\Sigma(r) = \Sigma_0
(r_0/r)^k$ at $r_1 < r < r_2$ (where $r_1 > a$) induces a prograde
apsidal precession \citep{1981Icar...47..234W}
\begin{equation}
{\dot \varpi}_{\rm disk} = 2 \pi n \left(\Sigma_0 a^2 \over M_\ast\right)
                           \left(r_0 \over a\right)^k
                           \left[W_k(r_1/a) - W_k(r_2/a)\right] ,
\end{equation}
where
\begin{equation}
W_k(r/a) = \sum_{\ell=1}^\infty
           {\ell (2\ell + 1) \over (2\ell + k - 1)}
           \left[(2\ell)! \over 2^{2\ell} (\ell!)^2\right]^2
           \left(a \over r\right)^{2\ell + k - 1} .
\end{equation}
If we approximate the azimuthally averaged density profile for $q_2 =
5.9 \times 10^{-3}$ in Fig. \ref{fig:sigma-multi1} as a flat profile
($k = 0$) with $\Sigma_0 = 7500\,{\rm g}\,{\rm cm}^{-2}$ between $r_1
= 0.6 \au$ and $r_2 = 1.2 \au$, we find ${\dot \varpi}_{{\rm disk},2}
\approx 0.\!\!^\circ7 \yr^{-1}$, which is in reasonable agreement with
the measured value. 
The disk-induced precession rate can be increased by increasing the
disk mass or decreasing the gap width $(r_1 - a)/a$, but the gap width
is determined mainly by the mass ratio $q$ and is not very sensitive
to either the viscosity parameter $\alpha$ or the disk thickness $H/r$
\citep[e.g.][]{1999MNRAS.303..696K, 1999ApJ...514..344B,
2004ApJ...612.1152V}. Increasing $\alpha$ and/or $H/r$ primarily
makes the gap shallower and not so deep.

We have performed three-body integrations similar to that in
Fig. \ref{fig:3body1}, but with additional apsidal precession
${\dot \varpi}_{{\rm disk},2} = p\,(0.35 \au/a_2)^{3/2} \,^\circ
\yr^{-1}$ and $p = 1$ and 10.
The much smaller disk-induced precession of the inner planet's orbit
was neglected.
Except for a larger libration amplitude for $\Delta\varpi$ in the $p =
10$ case, the evolution of $\theta_1$ and $\Delta\varpi$ is similar to
that shown in Fig. \ref{fig:3body1} for $p = 0$, and there is no
additional delay in the capture of $\Delta\varpi$.
Figure \ref{fig:3body2} shows the evolution of the eccentricities for
$p = 0$, 1, and 10.
At a given time, the additional positive apsidal precession results in an
increase in $e_1$ and a decrease in $e_2$, but the effect is small for
$e_1 < 0.31$ (the upper limit for GJ 876; Fig. \ref{fig:gj876}) even
for $p = 10$.
The additional apsidal precession and forced migration rates used in
the three-body integrations are proportional to $a_2^{-3/2}$ and hence
the mean motion $n_2$.
A decrease in the positive, disk-induced precession rate due to, e.g., disk
dispersal would cause the eccentricities to adjust to values
appropriate for the precession rate at a given time and approach the
$p = 0$ case without disk-induced precession.

The analytic theory developed by \citet{1981Icar...47....1Y} for the 2:1
resonances of Io and Europa and the Laplace resonance takes into
account the apsidal precession induced by the oblateness of Jupiter
(which is largest for the innermost satellite Io), and it can be
adapted to understand the effects of the disk-induced apsidal
precession (which is larger for the outer planet).
As the orbits converge toward the 2:1 mean-motion commensurability
(i.e., as $2 n_2 - n_1 < 0$ increases), the resonance $\theta_1 =
2\lambda_2 - \lambda_1 - \varpi_1$ would be encountered before the
resonance $\theta_2 = 2\lambda_2 - \lambda_1 - \varpi_2$ if the
apsidal precession is dominated by ${\dot \varpi}_{{\rm disk},j}$
(since ${\dot \theta}_j \approx 2 n_2 - n_1 - {\dot \varpi}_j$ and
${\dot \varpi}_{{\rm disk},2} \gg {\dot \varpi}_{{\rm disk},1} > 0$).
However, because the resonance-induced retrograde apsidal precession
is proportional to $1/e_j$ and much larger in magnitude than the
disk-induced prograde precession for small $e_j$, $\theta_1$ 
and $\theta_2$ (and hence $\Delta\varpi$) can be captured into
libration in a sequence that differs little in order or timing from
the case where there is no disk-induced precession.
When we include ${\dot \varpi}_{{\rm disk},j}$ in
Eqs. (\ref{eq:dw1dt}) and (\ref{eq:dw2dt}) for small $e_j$, stable
retrograde precessions of both orbits still require $\langle \theta_1
\rangle = 0^\circ$ and $\langle \theta_2 \rangle = 180^\circ$, and the
requirement that the orbits precess at the same rate on average
($\langle {\dot \varpi}_1 \rangle = \langle {\dot \varpi}_2 \rangle$)
implies the following relationship between the forced eccentricities:
\begin{eqnarray}
e_2/e_1 &=& q_1 n_2 C_2 / [- \beta q_2 n_1 C_1 +
              e_1 ({\dot\varpi}_{{\rm sec},2} - {\dot\varpi}_{{\rm sec},1})
\nonumber \\
        & & + e_1 ({\dot\varpi}_{{\rm disk},2} - {\dot\varpi}_{{\rm disk},1})
] .
\label{eq:e2e1}
\end{eqnarray}
Since ${\dot \varpi}_{{\rm disk},2} - {\dot \varpi}_{{\rm disk},1} >
0$, the disk-induced apsidal precession reduces $e_2/e_1$.
The decrease in $e_2$ and increase in $e_1$ seen in
Fig. \ref{fig:3body2} are consistent with this trend, but it should be
noted that the 2:1 resonance configurations with $M_1/M_2 \approx 0.3$
and $e_1 \ga 0.04$ are no longer dominated by the lowest order
resonant terms and those with $e_1 \ga 0.1$ also have $\theta_2$ and
$\Delta\varpi$ librating about $0^\circ$ instead of $180^\circ$ (see
above).
Although $e_2/e_1$ is fixed by the resonance conditions, a simple
expression like Eq. (\ref{eq:e2e1}) does not exist for the larger
eccentricities.

\begin{figure}[ht]
\begin{center}
\resizebox{0.98\linewidth}{!}{%
\includegraphics{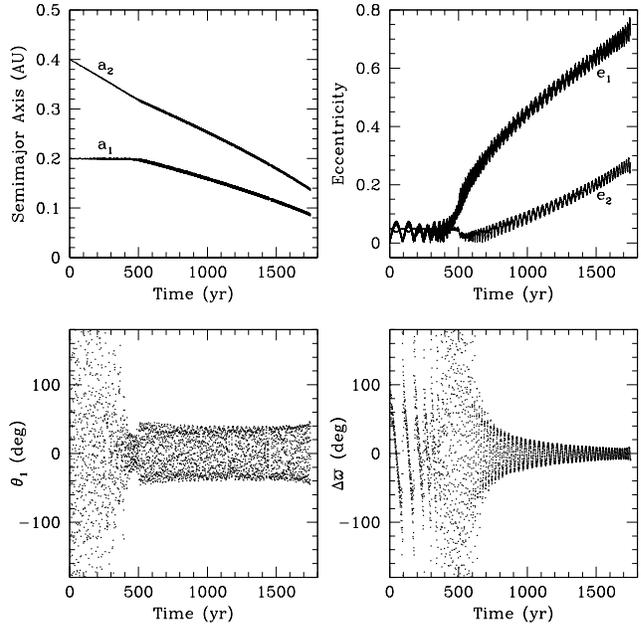}}
\end{center}
  \caption{
  Same as Fig. \ref{fig:3body1}, but for the three-body model with the
  following initial conditions: $e_1 = 0.01$, $e_2 = 0.05$, and the
  orbits are antialigned, with the inner planet at periapse and the
  outer planet at apoapse.
  }
   \label{fig:3body3}
\end{figure}

\subsection{Initial Eccentricities}
The isothermal model h2 (Fig. \ref{fig:g33h2}) shows large initial
eccentricity variations induced by the eccentric disk, with both $e_1$
and $e_2$ around $0.05$.
The large eccentricities (in particular $e_2$) mean that the planets
cannot be captured into the 2:1 resonance configuration with
$\theta_1$ librating about $0^\circ$ and $\theta_2$ librating about
$180^\circ$ (see Fig. \ref{fig:3body1}).
In Fig. \ref{fig:3body3} we show the results of a three-body
integration similar to the baseline model in Fig. \ref{fig:3body1},
but with the following initial conditions: $e_1 = 0.01$, $e_2 = 0.05$,
and the orbits are antialigned, with the inner planet at periapse and
the outer planet at apoapse.
These initial conditions were chosen so that the eccentricity
variations are similar to those in the model h2 when $a_2 \approx 0.35
\au$.
The larger value of $e_1$ has prevented the initial libration of
$\theta_1$ that is seen in Fig. \ref{fig:3body1}.
But $\theta_1$ is captured into libration about $0^\circ$ at about the
time ($t \approx 300 \yr$) when $a_2 \approx 0.35\au$, and its
libration amplitude is similar to that in model h2.
There is also a delay in the capture of $\Delta\varpi$ (and $\theta_2$)
into libration to $t \approx 600 \yr$ and, as expected, $\Delta\varpi$
is captured directly into libration about $0^\circ$. 
The libration amplitude of $\Delta\varpi$ is smaller than that in
model h2, possibly because of continuing interaction with an eccentric
disk in the latter case.
The relative timing of the captures can have no effect on the
necessity of large eccentricity damping if there is extensive
migration within the resonances, since the ratio of the eccentricities
is determined uniquely by the resonance conditions once all the
resonance variables are librating.

\section{Conclusion}
\label{sec:conclusion}
We have modeled the evolution of the \GJ\ system by performing
two-dimensional hydrodynamical simulations of a disk with embedded planets.
We confirm previous work showing that interactions between the outer
planet and a gas disk can drive two initially non-resonant planets
into resonance, and in particular we have shown that the induced migration
of the GJ 876 planets results in the capture of the 2:1 resonance variables,
$\theta_1,\,\theta_2$ and $\Delta\varpi$ into the observed libration
about $0^\circ$.
The precession of the outer planet's periapse longitude induced by the
disk was shown not to increase the delay of capture of $\theta_2$ into
resonance, but a high value of $e_2$ induced by an eccentric disk did
increase the delay.  Later capture of $\theta_2$ has no effect on the
eccentricity damping problem as the ratio of the forced eccentricities
is fixed once both $\theta_1$ and $\theta_2$ are librating.

The ``isothermal'' simulations with fixed disk temperature structure
have shown that more massive planets with a planet-to-star mass ratio
of $q \ge 5.2 \times 10^{-3}$ are able to perturb the disk
sufficiently to make it eccentric (even if the planetary orbit is
circular).
This effect may be caused by an eccentric instability driven
by an interaction of the $m=2$ mode at the outer eccentric 3:1 
Lindblad resonance with a slightly eccentric disk (see 
\citealt{2001A&A...366..263P}).
For smaller masses in the isothermal models and for all masses in the
radiative models, the planet-disk interaction produces the typical gap
and spiral arms, but the disk remains otherwise circular.

The simulations also confirm that for non-eccentric,
circular disks the eccentricity growth of the outer
planet is suppressed until $\theta_2$ is locked into libration about
$0^\circ$ (see models h4, m1, m2). However, $\theta_2$ and
$\Delta\varpi$ are always eventually captured into libration,
and the subsequent increase in the eccentricities of both planets past
the observational upper limits in \GJ\ ($e_1 \approx 0.31$, $e_2
\approx 0.05$) occurs on a time scale
shorter than the expected viscous time scale of the disk. The final
eccentricities of both planets are then substantially larger than
those seen in \GJ, a result which was found already in
previous hydrodynamic \citep{2003CeMDA..87...53P,2004A&A...414..735K} and three-body
simulations \citep[e.g.][]{2002ApJ...567..596L}.

In addition to causing an increased delay in the capture of $\theta_2$
and $\Delta\varpi$, eccentric disks (models h2, h9) lead to
significantly larger libration amplitudes for $\Delta\varpi$ ($\approx
50\degx$) than for the circular disk case.
Interestingly, it appears that in the case of \GJ\ the libration
of $\Delta\varpi$ is in fact larger ($\approx 34\degx$ for $i =
90^\circ$) than that of $\theta_1$ ($\approx 7\degx$ for $i =
90^\circ$), as shown recently by \citet{2004astro-ph0407441}.

\citet{2002ApJ...567..596L} have shown that if sufficient eccentricity
damping ($\dot e_2/e_2 \approx 100 \,\dot a_2/a_2$) is applied during
migration, the observed configuration of \GJ\ can be maintained for
arbitrary migration times.
We have updated the amount of eccentricity damping required using the
updated dynamical fits by \citet{2004astro-ph0407441}.
For the best fits with coplanar inclination $i \ga 35^\circ$, $e_1 \la
0.31$, $e_2 \la 0.05$, and $\dot e_2/e_2$ must exceed $\approx 40\,
\dot a_2/a_2$.
However, no such rapid eccentricity damping mechanism is presently known.
The hydrodynamic simulations typically give comparable time
scales for eccentricity damping and semi-major axis decrease.
Some current theories of
planet-disk interactions indicate that eccentricity {\em driving}
should occur \citep{2003ApJ...587..398O, 2003ApJ...585.1024G}. 
Such eccentricity behavior would not be consistent with the observed
state of \GJ\ if there were extensive migration after capture into the
resonances. 

If the disk is removed soon after the planets are captured into
resonance, as in Figs. \ref{fig:g33k3b} to \ref{fig:g33m2}, the driving
disappears and the final eccentricities are smaller, like
those observed.
However, the disk must vanish before the post-capture orbits shrink
by $\sim 10\%$, similar to the result found in 3-body integrations by
\cite{2002ApJ...567..596L}. 
We have used in the present simulations a high viscosity and high surface
density to increase the migration rates, and to be able to perform the 
simulations in a reasonable amount of computer time. In turn this
implies a need for 
rapid disk dispersal to halt migration.
If more realistic values were used for the viscosity
and the disk mass, the necessary disk
dispersal time scale could be much more extended.
What appears to be somewhat fine tuned in the present simulations
would be perhaps more reasonable. 
Our scenario only requires that capture occurred in the final formation
phase of \GJ, and that the planets did not migrate over a very
long distance while locked in the resonances.

Additional effects that might influence the eccentricity damping 
in these type of simulations are possibly three-dimensional.
Fully 3D-simulations of circular planets on a fixed orbit 
\citep{2003ApJ...586..540D, 2003MNRAS.341..213B}
have shown that the spiral structures are not so clear and
material may enter the Roche lobe from above the midplane, an effect
which may alter the eccentricity damping properties.
The inclusion of magnetic fields might lead to a magnetic coupling
of the planetary field with that of the disk, which will have an
influence on the eccentricity evolution. But if the consideration of these
and perhaps other processes fail to produce sufficient
eccentricity damping, the elimination of the disk shortly after the
capture of the GJ 876 planets into the 2:1 resonances is a possible,
albeit perhaps unsatisfying means of accounting for the observed
low-eccentricity configuration.

\begin{acknowledgements}
We would like to thank Greg Laughlin for stimulating discussions
during the course of this project and for furnishing the details of
the dynamical fits by \citet{2004astro-ph0407441}.
We also thank D. N. C. Lin and
M. Nagasawa for informative discussions. The work was sponsored by the
KITP Program ``Planet Formation: Terrestrial and Extra-Solar''
held in Santa Barbara from January 2004 until March 2004. We thank
the organizers for providing a pleasant and very stimulating
atmosphere. 
This research was supported in part by the National Science Foundation
under Grant No. PHY99-0794 and by NASA grants NAG5-11666 and
NAG5-13149.
\end{acknowledgements}

\bibliographystyle{aa}
\bibliography{2656lit}
\end{document}